\documentclass[12pt]{article}
\def\id{\protect{{1 \kern-.28em {\rm l}}}}

\pdfoutput=1
\usepackage{putex}
\usepackage{feyn}
\usepackage[vcentermath]{youngtab}

\usepackage{graphicx}
\usepackage{epstopdf}
\usepackage{enumerate}
\usepackage{cite}
\usepackage{tensor}
\usepackage{slashed}
\usepackage{feynmf}

\usepackage{hyperref}

\def\be{\begin{equation}}
\def\ee{\end{equation}}
\def\bea{\begin{eqnarray}}
\def\eea{\end{eqnarray}}

\makeatletter
\renewcommand\section{\@startsection {section}{1}{\z@}%
                                   {-3.5ex \@plus -1ex \@minus -.2ex}%
                                   {2.3ex \@plus.2ex}%
                                   {\normalfont\large\bfseries}}
\renewcommand\subsection{\@startsection{subsection}{2}{\z@}%
                                   {-3.25ex\@plus -1ex \@minus -.2ex}%
                                   {1.5ex \@plus .2ex}%
                                   {\normalfont\normalsize\bfseries}}
				   			
\def \foot {\footnote}

\def \tr {{\rm tr}}
\def \ha {{1 \over 2}}
\def \td {\tilde}
\def \ci{\cite}
\def \N {{\mathcal N}}

\def \L {\Lambda}

\def\z{\zeta}

\def\a{\alpha}
\def\b{\beta}

\def \del{\partial}
\def \a {\alpha}

\def\z{\zeta}

\def\ov{\over}

\def\b{\beta}
\def\l {\lambda}

\def \k {\kappa}
\def\foot{\footnote}

\def\det {\hbox{det}}
\def \ci {\cite}

\def \l  {\lambda}

\def \N {{\mathcal N}}

\def \td {\tilde}

\def \D {\Delta}
\def \N {{\mathcal N}}

\def \la {\label}

\def \l {\lambda}
\def\foot{\footnote}

\newcommand{\rf}[1]{(\ref{#1})}
\def \ov {\over}

\def\N{{\cal N}}

\def \ha{{1\ov 2}}
\def \r {\rho}

\def \no {\nonumber}

\def \del {\partial}

\def \om {\omega}

\def \la {\label}
\def\foot{\footnote}

 \def \r {\rho}

\def \ov {\over}

\def \varpi {{\rm w}}

\def \ep {\epsilon}

\def \ed {\end{document}}

\def \te {\textstyle}

\def \ha {{{\textstyle{1 \ov2}}}}
\def \fo {{\textstyle{1 \ov4}}}

\def \D  {\Delta }

 \def \eqref  {\rf}
\def \ha {{\textstyle {1 \ov 2}}}

\renewcommand{\theequation}{1.\arabic{equation}}
\def \be {\bea}
\def \ee {\eea}

\def \pe {\perp} \def \tri {{\te { 3 \ov 2}}}
\def \iffa {\iffalse}

  \def \De {\Delta}

 \def \kk {{\rm k}}

\def \k {\kappa}

 \def \L {\Lambda}
\def \D {{\rm D}}

\def  \iffa {\iffalse}
\def \tri {\triv}

\def \ep {\epsilon}

\def \text { }
\def \pe {\perp}

\begin{document}


\overfullrule=0pt
\parskip=2pt
\parindent=12pt
\headheight=0in \headsep=0in \topmargin=0in \oddsidemargin=0in

\vspace{ -3cm}
\thispagestyle{empty}
\vspace{-1cm}

\date{\today}

 \def \G {\Gamma}

\preprint{PUTP-2460\\ Imperial-TP-AT-2014-01}

\institution{PU}{Department of Physics, Princeton University, Princeton, NJ 08544}
\institution{PCTS}{Princeton Center for Theoretical Science, Princeton University, Princeton, NJ 08544}
\institution{IMP}{Blackett Laboratory, Imperial College, London SW7 2AZ, U.K.}

\title{
Partition Functions and Casimir Energies\\
in Higher Spin AdS$_{d+1}$/CFT$_d$
}

\authors{Simone Giombi,\worksat{\PU} Igor R.~Klebanov\worksat{\PU,\PCTS} and Arkady A. Tseytlin\worksat{\IMP}\footnote{Also at Lebedev  Institute, Moscow.}
}

\abstract{
Recently, the one-loop free energy of higher spin (HS) theories
in Euclidean AdS$_{d+1}$ was calculated and matched with the order $N^0$ term in the free energy of the large $ N$ ``vectorial"
scalar CFT on the $S^d$ boundary. Here we extend this matching to the boundary theory defined on $S^1 \times S^{d-1}$,
where the length of $S^1$ may be interpreted as the inverse temperature.
It has been shown that the  large $N$ limit  of the
partition function on $S^1\times S^2$ in the $U(N)$ singlet sector of the CFT of $N$ free complex scalars
matches the one-loop thermal partition function of the Vasiliev theory in AdS$_4$,
while in the $O(N)$ singlet sector of the CFT of $N$ real scalars it matches the minimal theory containing even spins only.
We extend this matching to all  dimensions $d$.
We also calculate partition functions for the singlet sectors of free fermion CFT's in various dimensions and match them with
appropriately defined higher spin theories, which for $d>3$ contain massless gauge fields with mixed symmetry.
In the zero-temperature case $R \times S^{d-1}$ we calculate the Casimir energy in the scalar or fermionic CFT and match it with the one-loop
correction in the global AdS$_{d+1}$. For any odd-dimensional CFT the Casimir energy must vanish on general grounds, and we show that the HS duals
obey this. In the $U(N)$ symmetric case, we exhibit the vanishing of the regularized 1-loop Casimir energy of the
dual HS theory  in AdS$_{d+1}$.
 In the minimal HS  theory  the vacuum energy vanishes for odd $d$  while for even $d$ it is equal to the
 Casimir energy  of a single  conformal scalar in $R \times S^{d-1}$ which is again consistent with
 AdS/CFT, provided the  minimal HS coupling constant is $\sim 1/(N-1)$.
 We demonstrate analogous results for singlet sectors of theories of $N$ Dirac or Majorana fermions. We also discuss extensions to
 CFT's containing $N_f$ flavors in the fundamental representation of $U(N)$ or $O(N)$.}

\maketitle

\tableofcontents

\def \D  {\hat\Delta}\def \dD {{\rm d}}
 \def \zr {{\z_{\rm R}}}
\def \tri  {{\te{3 \ov 2}}}
\def \N {{\cal N}}\def \K {{\rm K}}
 \def \AdS {{AdS}}
 \def \K {{\rm K}} \def \kk {{\rm k}}
  \def \De  {\Delta}
\def \zz {{\cal Z}}
\def \dd {{\rm d} }
 \def \zhs {\z_{\rm HS}}
   \def \zh {\z_{E}}

\def \rmr  {{\rm \ell}}
 \def \half {{1 \ov 2}}
\def \om  {\omega}

\setcounter{equation}{0}
\setcounter{footnote}{0}
\section{Introduction and summary}

The original AdS/CFT conjectures were made for conformal field theories of $N\times N$ matrices with extended supersymmetry \cite{Maldacena:1997re,Gubser:1998bc,Witten:1998qj}.
A few years later, a suggestion was made to study AdS/CFT correspondence for simpler field theories where dynamical fields are
in the fundamental representation of the $U(N)$ or $O(N)$ symmetry group \cite{Klebanov:2002ja}; for this reason, these theories are often called
``vectorial." In these cases, the supersymmetry is not necessary,
but it is important that there is an infinite tower of conserved or
nearly-conserved higher spin currents that are $U(N)$ or $O(N)$ singlets. Therefore, the dual theories in AdS must contain
the corresponding tower of massless higher-spin gauge fields \cite{Sundborg:2000wp}.
Theories of this kind have been extensively explored by Vasiliev and others \cite{Fradkin:1987ks,Vasiliev:1990en,Vasiliev:1992av,Vasiliev:1995dn,Prokushkin:1998bq,Vasiliev:1999ba, Vasiliev:2003ev, Bekaert:2005vh}.

The first explicit vectorial AdS/CFT conjectures were made for the higher spin theories in AdS$_4$. For the minimal type A theory with even spins only, the
conjectured duals were the free or critical $O(N)$ models, with $N$ real scalar fields in the fundamental representation. For the non-minimal type A
theory, where all integer spins are present, one instead needs to consider free or critical theories of $N$ complex scalar fields, restricted to the $U(N)$
singlet sector \cite{Klebanov:2002ja}. There also exist type B Vasiliev theories in AdS$_4$ where the bulk spin zero field is a pseudoscalar, rather than a scalar.
Such theories have been conjectured to be dual to the $U(N)$ or $O(N)$ singlet sector of the theory of $N$ Dirac or Majorana fermions \cite{Sezgin:2003pt,Leigh:2003gk}.

The basic evidence for the initial conjectures involved the matching of the spectra of currents and higher-spin gauge fields
\cite{Klebanov:2002ja,Sezgin:2003pt,Leigh:2003gk}.
A nice way of summarizing this agreement is to match the CFT partition function on $S^1\times S^2$ with the corresponding calculation in
AdS$_4$. This was carried out in \cite{sy}, and a simple explicit formula for the partition function of the $U(N)$ singlet scalar
 theory was obtained. A crucial ingredient in these
calculations is the imposition of the singlet constraint in the CFT of free scalar fields; this was accomplished by integrating over the holonomy of the
$U(N)$ gauge field around $S^1$ \cite{Sundborg:1999ue,aha,sh}. The resulting $U(N)$ singlet partition function then becomes the square of the character of the ``singleton''
representation of $SO(3,2)$, corresponding to the free conformal scalar in $d=3$. The CFT partition function may then be expanded in characters of the primary fields of spin $s$ and dimension $s+1$, which correspond to
partition functions of gauge fields in AdS$_4$.
Besides providing a nice test of the vectorial AdS$_4$/CFT$_3$ duality \cite{Klebanov:2002ja},
this may be viewed as a modern incarnation of much older ideas \cite{Flato:1978qz} (see also \cite{Kutasov:2000td,Polyakov:2001af, bia1, bia2, Vasiliev:2004cm, Barabanschikov:2005ri,dolan}).
The $d=3$ result of \cite{sy} was recently reproduced  and also extended to the $O(N)$ singlet sector of $N$ real scalars \ci{jev}
using the collective field approach \cite{Das:2003vw}.
We will further extend these results in several ways, thus obtaining new tests of the higher spin AdS/CFT dualities.

Additional evidence for the vectorial AdS$_{d+1}$/CFT$_d$ duality for $d=3$ has been found in
\cite{Giombi:2009wh,Giombi:2010vg,Giombi:2011ya,Maldacena:2011jn,Maldacena:2012sf,Colombo:2012jx,Didenko:2012tv,Gelfond:2013xt,Didenko:2013bj,gk}.
Furthermore, evidence has begun accumulating that it is valid for all $d$ \cite{Didenko:2012vh,gks}.
On the CFT side, we may consider $N$ complex or real scalar fields in $d$ dimensions and impose the $U(N)$ or $O(N)$ singlet constraint.\footnote{
Perhaps the constraint can implemented by coupling the free $N$-vector theory to an appropriate topological gauge theory, generalizing the idea of coupling to
Chern-Simons theory in $d=3$ \cite{Giombi:2011kc, Aharony:2011jz}. For the purposes of this paper, the details of how the singlet constraint is imposed do not seem to matter.}
The corresponding theories in AdS$_{d+1}$, which involve the tower of totally symmetric tensor spin $s$ gauge fields were formulated in \cite{Vasiliev:2003ev}.
For arbitrary $d$, we will study the partition function of such a theory in ``thermal" AdS space, which is asymptotic to $S^1\times S^{d-1}$ and
match it  with the singlet partition function of the free scalar theory on  $S^1\times S^{d-1}$.
This result provides
an elegant way of encoding the AdS/CFT matching of the spectra.
The thermal free energy on $S^1\times S^{d-1}$ includes a term linear in the inverse temperature which dominates in the zero temperature limit. This is related to the
Casimir energy of the CFT on $R\times S^{d-1}$, and may be computed by an appropriately regularized sum over the energy spectrum. We will compare this Casimir energy term on
the two sides of the duality.
In this case, the higher spin theory
 is defined
 in the global AdS$_{d+1}$, which is asymptotic to
$R\times S^{d-1}$. For all odd $d$ the Casimir energy must vanish; this is a completely
general fact about odd-dimensional CFT related to the
absence of anomalies (the theory on $R\times S^{d-1}$ may be obtained from that on $R^d$ via a Weyl transformation).
We check this vanishing on the higher spin side by using
an appropriate zeta-function regularization of the sum over spins in global AdS$_{d+1}$. 
The vanishing of Casimir energy in $d=3$ was perhaps
the reason why it was not emphasized in \cite{sy}. However, the vanishing in odd $d$ is not trivial from the AdS point of view because it involves summing over
the entire tower of spins. Truncation of the spectrum in AdS$_4$ to a few low spins, which is commonly performed in
``bottom-up" modeling, would generally spoil the cancelation of the Casimir energy. This would violate a possible exact duality to a CFT$_3$,
unless there is another reason for the bulk cancelation, such as supersymmetry (as in  \ci{ad,gn,iy}).

The comparison of the Casimir energies becomes even more interesting for even $d$, where they are not required to vanish.
For the ${\cal N}=4$ $SU(N)$ gauge theory in $d=4$, the ${\cal O}(N^2)$ term in the Casimir energy was reproduced early on using the stress-energy tensor calculation in
AdS$_5\times $S$^5$ \cite{Balasubramanian:1999re}. Due to the cancelation of the total derivative (``$D$-anomaly'') terms
in the full expression for the  trace  anomaly,
its Casimir energy is proportional to its $a$-anomaly coefficient \ci{herz,hua}.
Therefore, the exact AdS/CFT matching of Casimir energy in that case is guaranteed by the $a$-anomaly matching \cite{Henningson:1998gx}.
In the field theory, the exact result  is $a=N^2-1$,
and the  ${\cal O}(N^0)$ correction (i.e. the -1  shift)   has   been
studied using the one-loop correction in the type IIB supergravity on the  AdS$_5\times $S$^5$  background \cite{Mansfield:2002pa}.
More recently, additional progress has been made in calculating the ${\cal O}(N^0)$ correction to $a-c$ in various $d=4$ theories, where only contributions of short
supermultiplets in AdS$_5$ need to be included \cite{Ardehali:2013gra, Ardehali:2013xya}.

In non-supersymmetric theories, the Casimir energy is not simply proportional to $a$ due to the presence of the total derivative anomaly terms
\ci{herz,hua}.
This makes the comparison of Casimir energies a new check of the vectorial AdS/CFT conjectures, which is independent of the comparison of
$a$-anomalies carried out in \cite{gk, gks}. Unfortunately, due  to the  lack of information about the form of  the classical action,
in the higher-spin theories there is no known way to calculate the leading,  ${\cal O}(N)$,
terms in the sphere free energies or Casimir energies. So, as in \cite{gk,gks}, we will only compare the terms of order $N^0$.
In the non-minimal Vasiliev theory including all integer spins, we find that the regularized sum in AdS$_{d+1}$ vanishes,
 in line with the expectation
that there is no ${\cal O}(N^0)$ correction in the free
complex scalar theory on $R \times S^{d-1}$.
 However, in the minimal theory, which includes even spins only, the regularized
sum equals the Casimir energy of a real scalar field. These results are analogous to the recent findings in Euclidean  AdS$_{d+1}$, where the one-loop
correction for the minimal theory in AdS$_{d+1}$ was found to be equal to the free energy of a single real scalar on $S^d$ \cite{gk, gks}. The proposed interpretation of this result is that the bulk coupling constant in the minimal higher spin theory is $G_N\sim 1/(N-1)$, so that the tree level and one-loop terms can add to give the answer which is $N$
times the contribution of a free scalar field. Our new results for Casimir energies in all $d$ provide additional support for this interpretation.

In this paper we also study the vectorial fermionic models on $S^1\times S^{d-1}$ and match their partition functions and Casimir energies
 with the corresponding quantities in AdS$_{d+1}$.
 Such calculations are quite useful: for $d>3$ the dual higher spin theory in AdS$_{d+1}$ includes massless gauge fields in
mixed symmetry representations \cite{Anselmi:1998bh, Anselmi:1999bb,Vasiliev:2004cm,dolan,Alkalaev:2012rg}, in addition to the totally symmetric higher spin fields
found in the Vasiliev theories dual to the scalar field theories \cite{Vasiliev:2003ev}. The AdS spectrum dual to a fermionic model
depends sensitively both on the dimension $d$ and on what type
of fermions we consider: Dirac, Majorana or Weyl. These results suggest the existence of a variety of consistent interacting higher spin theories
that are dual to fermionic CFT's restricted to singlet sectors.


We start  in Section \ref{back} with a  brief  summary  of some standard relations  between  Casimir energies and  partition  functions,
and then in Section \ref{free-therm}  review the   expression  for the free energy of free conformal fields  in $S^1 \times S^{d-1}$.
In Section \ref{singlets} we compute  this free energy   for a large number $N$   of complex or real scalar or fermion
 fields  in the presence of
a singlet constraint. The latter   translates into  an  extra gaussian averaging   over the density of  $U(N)$ or $O(N)$  holonomy
 eigenvalues  that  leads to a  modification  of the effective  one-particle  partition function. 
 The  resulting  free  energy contains an order $N$   Casimir energy term as  well as an order $N^0$  term with
 non-trivial $\b$-dependence.   The scalar  free energies
   are matched  onto  the corresponding expressions  in the
  dual HS theories  in AdS$_{d+1}$ in Sections \ref{therm-HS}.
  Section \ref{free-fermions} contains a similar analysis of the vectorial fermionic  CFT's  in $d=2,3,4$ and of their higher spin duals.
  For each admissible type of fermion, we study the quantum numbers of the currents and corresponding gauge fields,
and demonstrate the AdS/CFT matching of the Casimir
energies and partition functions. In Section \ref{HS-Nf} we briefly discuss the HS duals of the CFT's containing $N_f$ fundamental flavors of
$U(N)$ or $O(N)$. In the large $N$ limit where $N_f$ is held fixed, we demonstrate the matching of partition functions and Casimir energies with
the field theory results.

\renewcommand{\theequation}{2.\arabic{equation}}
 \setcounter{equation}{0}
\section{General background}
\label{back}

Given a CFT in  $d$ dimensions,  in the standard radial quantization  picture  its states  may  be described as
eigenstates of the Hamiltonian
on $R_t \times S^{d-1}$.
Given a  set of states  and ignoring interactions
one may then consider, e.g.,  the corresponding  Casimir energy  and construct the finite temperature partition function.
The same quantities may be computed  also  on the dual AdS$_{d+1}$ side
as the vacuum energy
in the global AdS or  as  the 1-loop partition function
on a thermal quotient of AdS, i.e. on Euclidean AdS   with boundary $S^1_\beta  \times  S^{d-1}$.

Let us summarize some   standard relations (see, e.g., \ci{gpp}).
Given the spectrum   of a Hamiltonian $H$ (with  eigenvalues
 $\om_n$  and degeneracies $\dd_n$  where  $n$ is  a multi-index)
   one may consider  the     ``energy''   zeta-function
\be
\z_E (z) = \tr\,   H^{-z} =  \sum_n { \dd_n\, \om^{-z}_n}  \ , \la{01}
\ee
  so that  the Casimir  or vacuum energy   is given by (for fermions one
needs to  add a minus sign)
\be
E_c= \ha \sum_n { \dd_n\, \om _n}  = \ha \z_E (-1)  \ . \la{02}
\ee
One may also define  the  one-particle or canonical  partition function\foot{For simplicity we shall ignore possible chemical potentials.}
\be
\zz(\b) = \tr\, e^{-\b H} =  \sum_n  \dd_n\, e^{-\b \om_n}   \ . \la{03} \ee
It is   related  to $\z_E (z) $ by the Mellin transform
\be
\z_E (z) ={1 \ov \G(z)} \int^\infty_0 d\b\, \b^{z-1} \zz(\b)  \ ,  \la{04} \ee
i.e.  the two functions contain an equivalent amount of information about the spectrum. This
is the same as the usual   relation  between a spectral zeta function  for an  operator  $\Delta$ (here $\Delta =H$)
and its  heat kernel (here $\zz(\b)  =  K(\tau ) =\tr \, e^{-\tau  \Delta} $  with $\b$ playing role of $\tau$).
Note also  that a special   case of \rf{04}    is the   integral representation for the standard Hurwitz zeta function
\be
\z(z,a) =\sum_{k=0}^\infty (k+ a)^{-z}=  {1 \ov \G(z)} \int^\infty_0 d\b\, \b^{z-1}  { e^{- a\b} \ov 1-  e^{-\b} }
\ .
\la{041} \ee
The  multi-particle or grand canonical  partition function   which for bosons is
\be  \ln Z(\b) =  \tr \ln { \big(1 - e^{-\b H}}\big)^{-1}  = - \sum_n d_n \ln  (  1 - e^{-\b \om_n})
 \ , \la{05} \ee
is then directly related to the one-particle one \rf{03}, with the free energy given by
 \be
F_\b =-\ln Z(\b) = -\sum_{m=1}^\infty  { 1 \ov m}  \zz(m \b)  \ . \la{06}  \ee
For fermions
\be  \ln Z(\b) =  \tr \ln\big(  {   1 + e^{-\b H}}\big)  =  \sum_n d_n \ln  (  1 +  e^{-\b \om_n})
 \ , \la{05f} \ee
is then directly related to the one-particle one \rf{03}, with the free energy given by
 \be
F_\b =-\ln Z(\b) =  \sum_{m=1}^\infty  { (-1)^m \ov m}  \zz(m \b)  \ . \la{06f}  \ee
 Thus the knowledge of the one-particle  partition function $\zz(\b)$
  determines   the thermodynamic
partition function \rf{06}  as well as  the Casimir energy  (see \rf{02},\rf{04}).

\renewcommand{\theequation}{3.\arabic{equation}}
 \setcounter{equation}{0}
\section{Partition functions for free CFT's on  $S^1 \times S^{d-1}$}
\label{free-therm}
\def  \dq   {{q}}

Let   us first consider the  partition function  of a free conformally coupled  scalar
\be \la{11}
F=- \ln Z = \ha \ln \det \Delta_0  \ , \ \ \ \ \ \ \ \ \  \ \ \
 \Delta_0 = -\nabla^2 + {\te {d-2 \ov 4 (d-1)} }R \ , \ee
in (Euclidean)  $M^d = R \times S^{d-1} $
and $M^d_\b= S^1  \times S^{d-1} $  where $\b $ is the length of $S^1$.\foot{We shall  often
set the radius  $\rmr$ of $S^{d-1} $
to 1;  dependence on it can be restored by   rescaling $\b \to {\b \ov \rmr}$, etc.}
We  shall assume   the length of time direction in  $R\times S^{d-1}$   to be  regularized as  $\int dt= \b \to \infty$, so that
the   first case  may be    viewed as  the ``zero-temperature'' ($\b^{-1} \to  0$)  limit of the second.
In general,  in    $R \times S^{d-1} $   one  finds (see, e.g., \ci{Dowker:1978md,Dowker:1983ci,Blau:1988kv,coc})
\be \la{12}
F= F_\infty + F_c  \ , \ \ \ \ \ \ \   \ \ \ \ \ \ \ \ \ \ \   F_\infty =   a_d  \ln \L   \ , \ \ \ \ \ \ \ \
     F_c= \b    E_c \ , \    \ee
where  we separated  an a priori   possible  logarithmically divergent term
from  the  vacuum (Casimir)  energy    $E_c$  of a conformal scalar
 in the static Einstein  universe    $R \times S^{d-1} $.
 The  logarithmically divergent part (with $\L$ standing for the  product of a  UV cutoff with the scale $\ell$)
 is proportional to the conformal anomaly coefficient  $a_d$
 which vanishes for odd $d$.\foot{In \rf{12}  we include the volume factor $\sim \b$ into $a_d$.}
    In  the present case  it vanishes also   for even  $d$ as
 for a  conformally-coupled scalar  in a conformally-flat space  it is proportional to the Euler number density but  the latter  vanishes
 for both $R \times S^{d-1}$ and $S^1 \times S^{d-1}$, i.e.
 \be  \la{123}
 a_d=0 \ . \ee
The scalar curvature  of  $S^{d-1}$ is
$R=(d-1) (d-2)$
 so that the operator in \rf{11}  is
 $\Delta_0 =  - \del_t^2 -     \nabla^2_{S^{d-1}}  + {1 \ov 4 } (d-2)^2 $.
 Since  the eigenvalues  and their degeneracies for a Laplacian $-\nabla^2$ on a sphere  of dimension $d-1$
 are
 \be  \l_n ( S^{d-1}) = n (n + d-2) \ , \ \ \ \ \ \ \ \ \
{\te    \dd_n (S^{d-1})  =  { n + d-1 \choose d-1} -   { n + d-3  \choose d-1} =   (2n + d-2) { (n+ d-3)! \ov  (d-2)! n!} } \ \la{c0c} ,
  \ee
   the eigenvalues of $\Delta_0$ on $R\times S^{d-1} $  are  
 $\l_{w,n}= w^2 +   \omega^2_n$,\   where  $ \omega_n=  n + \ha (d-2)  $,\   $n=0,1,2,...$ and  $w\in (-\infty, \infty)$. 
  There is no  zero mode for $d > 2$.
 Integrating  $ \int^{\infty}_{-\infty}  {d w\ov 2 \pi}  \ln ( w^2 +  \omega^2_n) $ over $w$  leads as usual to $F_c= \b E_c$ with
   \be \la{a1}
  E_c = \ha \sum_{n=0}^\infty   \dd_n\,  \omega_n =  \sum^\infty _{n=0}  \te { { ( n + d-3)! \ov (d-2)! n!} \big[ n + \ha (d-2) \big]^2} \ .
  \ee
  This  is  finite if defined using the zeta-function regularization (see, e.g., \ci{coc,oz1,oz2,byt}),
  i.e. by starting as in \rf{01},\rf{02} with  $\om_n$ as energy eigenvalues, one first computes
  $\z_E(z) \equiv   \sum_{n=0}^\infty   \dd_n\,  \omega^{-z}_n$, and then analytically continues to $z=-1$,
  $E_c = \ha \z_E(-1)$.
  \foot{Note that  if we consider  the   conformal scalar  operator  on $S^{d-1}$, i.e.
  $\Delta_{0c}(S^{d-1})  =   -     \nabla^2_{S^{d-1}}  + {1 \ov 4 } (d-2)^2 $ then its eigenvalues are
  $ \l_{n} =  \omega^2_n$ with $  \omega_n= n + \ha ( d-2)$
   and the corresponding  spectral zeta function is
  $\zeta_{\Delta_{0c}(S^{d-1})}(z)  = \sum_{n=0}^\infty   d_n \l_{n}^{-z}  =  \z_E(2z)$.
  In particular, $E_c = \ha \z_E(-1) = \ha  \zeta_{\Delta_{0c}(S^{d-1})}(- \ha )$  \ci{coc}.
  Since   the natural spectral   parameter is $ n + \ha ( d-2)$ the  vacuum  energy $E_c$ can be expressed in terms of
  the corresponding Hurwitz  zeta functions. It can be also   computed using an exponential regularization
  $e^{-\ep [  n + {1\ov 2}  ( d-2)]}$ (dropping all singular terms).
  }

  In the case  of   $M^d_\b= S^1  \times S^{d-1} $
  the eigenvalues of $\Delta_0  $ are
  \be \la{111} \l_{k,n}= ({2 \pi k\ov \b} )^2  +   \omega^2_n \ , \ \ \ \ \ \ \ \  \omega_n =n + \ha (d-2) \ , \ \ \ \ \
  k=0, \pm 1, \pm 2, ... \ , \ \ n=0,1,2,... \ee
    One may define the  spectral   zeta function
    \be \la{13}
    \zeta_{\Delta_0}(z) = \sum_{k=-\infty}^\infty \sum_{n=0}^\infty  \dd_n \l^{-z} _{k,n}
  \ee
  in terms of which   we have for $F$ in \rf{11} (see, e.g., \ci{Allen:1986qi,Fursaev:2001yu})
  \be \la{14}
 && F= -  \zeta_{\Delta_0}(0) \ln \L   - \ha   \zeta'_{\Delta_0}(0) =   F_\infty  + F_c   + F_\b   \ , \\
 &&  F_\infty  =   a_d   \ln \L    \ , \ \ \ \ \ \ \ \ \      F_c= \b   E_c =   \ha \b \sum_{n=0}^\infty   \dd_n\,  \omega_n   \ ,\la{15}  \\
 &&   F_\b = \sum_{n=0}^\infty  \dd_n \ln (1 - e^{-\b \omega_n})  \ . \la{16}
  \ee
  Here   again $a_d=0$  for a  conformally-coupled scalar on $S^1_\b  \times S^{d-1} $  in  any $d$
  and
  $F_c$  is the same  Casimir energy  part  with $E_c$   given by  \rf{a1}.
  The non-trivial part of the  free energy $F_\b$ vanishes in the limit $\b \to \infty$
  when \rf{14} reduces to \rf{12}.

    Using the standard Riemann $\z$-function regularization  (with $\zeta (-2k) =0,\  \zeta (-2k-1) \not=0$)
    one finds for
     the Casimir energy  for $d >2$    (see, e.g., \ci{coc,oz1,oz2})
     \be
     && E_c  = \sum^{ [{d-3 \ov 2}] }_{q=0}   c_q\,  \z ( 2 q+ 1-d )  \  , \la{vvv}\\
   &&  d={\rm odd} \geq 3 :   \ \ \   E_c =0 \ ; \ \ \ \ \ \ \ \
    d={\rm even} \geq 4 : \ \ \
     E_c  = \sum^{ {1 \ov 2} d - 2 }_{q=0}   c_q\,  \z ( 2 q + 1-d )  \ , \la{1va} \ee
  where $c_q$   are  rational coefficients. Thus $E_c$ is non-vanishing in even $d$ and can be expressed in terms of the
Bernoulli numbers  (see  also below).

We conclude that
\be  d={\rm odd} \geq 3 :\ \ \ \ \ \  F=  F_\b \ ; \ \ \ \ \ \ \ \ \ \ \  d={\rm even} \geq 4 :\ \ \ \ \ \
F= \b E_c  + F_\b  \ , \la{fef} \ee
where $F_\b$  in  \rf{16} has  the following
   explicit form (cf. \rf{03},\rf{06})
\be \la{17}
&&   F_\b = - \sum_{m=1}^\infty   { 1 \ov m} \zz_0 (m \b) \ , \ \\
&&\zz_0 (\b) = \sum_{n=0}^\infty  \dd_n \,  e^{-\b [ n + \half(d-2) ]} \ =  {q^{ {d-2 \ov 2} } ( 1 + q)
 \ov ( 1 - q)^{d-1} } = {q^{ {d-2 \ov 2} } ( 1 - q^2)  \ov ( 1 - q)^{d} } \ , \ \ \ \ \  \ \ \ \   q\equiv e^{-\b} \ . \la{18}  \ee
The one-particle partition function $\zz_0(\beta)$ corresponds to the character of the free scalar (Dirac singleton) representation of the conformal
group $SO(d,2)$, see for instance \cite{dolan}.

Let  us add a few details about the explicit values  of the   Casimir energy \rf{a1},\rf{1va}.
For odd $d=2k+1$  one  can rewrite \rf{a1}    as
  $ E^{(2k+1)}_c =    \sum^\infty _{n=0}
    \sum_{m=1}^{k}  c_m   \big(n+ \ha\big)^{2m}$.
    For example,  $E^{(3)}_c = \sum^\infty _{n=0} \big(n+ \ha\big)^{2}$, etc.
  Since  $ \sum^\infty _{n=0}   \big(n+ \ha\big)^{2m} =  \zeta(-2m,\ha)=0$
    one concludes that  the  Casimir energy vanishes  for all odd $d=2k+1$.\foot{This agrees with
     the vanishing of conformal anomaly in odd  dimensional   space as there is a relation
  between a  combination of Euler density and total derivative  conformal anomaly coefficients,  and the Casimir energy of a conformal scalar
  (or more general CFT),
  as discussed in \ci{coc,herz} and references there.}
 For even $d=2k$ one  finds 
 \be \la{a3}
 E^{(2k)}_c =    \sum^\infty _{n=1}     \sum^{ k-1}_{m=1}   \td c_m    n^{2m+1}  \ ,   \ee
 which  is equivalent to the expression in \rf{1va} ($\td c_m = c_{-m - 1 + k}$).
 For example,
 \be E^{(2)}_c  &=&  \sum^\infty _{n=1}   n =\te  \zeta(-1) = -{1 \ov 12} \ , \label{Casexample}\\
 E^{(4)}_c &=& \ha   \sum^\infty _{n=1}   n^3  =\te \ha  \zeta(-3) = {1 \ov 240} \ , \nonumber \\
 E^{(10)}_c &=& \te
 {1 \over 40320}   \big[ \z(-9)  -  14 \z(-7)  + 49 \z(-5)  - 36 \z(-3)\big]  =-{317\over 22809600}  \ . \nonumber
 \ee
The same results    can be reproduced  by
introducing  a cutoff ($\ep \to 0$)  with the spectral   parameter  in the exponent
  $e^{-\ep  [n+ {1 \ov 2}  (d-2) ]}$   in   the sum  in \rf{a1}
  and dropping all singular terms in the resulting expression
 \be \la{aaa1}
E_c(\ep)  = \ha { e^{-{1 \ov 2} \ep d}  \ov (1 - e^{-\ep})^d}  \Big[d + ( d-2) \cosh\ep \Big]  \ .  \ee
Let us also note that an equivalent expression for the Casimir energy (\ref{a1}) may be also obtained
by a Mellin transform of the
one-particle partition function (\ref{18}), see eqs. (\ref{02}),(\ref{04}). This gives
\be\la{zet}
E_c = \ha \zeta_E(-1)\,,\qquad\qquad  \zeta_E(z) = \sum_{n=0}^{\infty} {\te { n + d -2 \choose d -2} }
\te
\left[(n+\frac{d}{2}-1)^{-z}+(n+\frac{d}{2})^{-z}\right]\,.
\ee
In this expression, the parameter $z$ provides a natural analytic regulator, as one can can perform the sum in terms of the Hurwitz zeta function and then analytically continue to $z=-1$. One may verify that this expression for $E_c$ gives the same values quoted above.  In particular, one can see directly the relation to \rf{a1}   by rearranging  the expression for
$ \zeta_E(z)$   as follows (cf. \rf{c0c})\foot{One is   to   change the summation variable  in the  second term $n \to  n'-1$ with $n'$ running now from 1,  and then observe  that the $n'=0$ term in the sum  can be added without altering  the result.}
\be
\zeta_E(z) = \sum_{n=0}^{\infty}  \dd_n\te  \left[n+ \ha ({d-2})\right]^{-z}\,, \ \ \ \ \
\dd_n =  \te { n + d -2 \choose d -2}  +  { n + d -3 \choose d -2}  =
 2  \big[ n +\ha  {(d-2)}\big] \frac{(n+d-3)!}{(d-2)! \ n! } \
\la{re}\ee

It is straightforward to generalize the above  analysis   to the case of  free  complex or real fermion theories.
First, for a single  massless Dirac fermion  in $S^1 \times S^{d-1}$   the free energy \rf{06f}  is given by the following
analog of the free conformal scalar expressions in \rf{17},\rf{18}:
\be \la{17f}
   F_\b =  \sum_{m=1}^\infty   { (-1)^m \ov m} \zz _{_{\rm F}} 
   (m \b) \ , \ \ \ \ \ \ \ \ \ \ \  \ \ \ \ \ \
\zz_ {_{\rm F}}
=  2 { 2^{[{d\ov 2}]}\  q^{ {d-1 \ov 2} }   \ov ( 1 - q)^{d-1} } \ ,  \ \ \ \ \ \   q= e^{-\b} \ .  \ee
 The result  for a real (Majorana) fermion  or Weyl   fermion  is   half of that,
\be
\zz_{\half}  (\b)= 
 { 2^{[{d \ov 2}]}\ q^{ {d-1 \ov 2} }   \ov ( 1 - q)^{d-1} } \ . \la{maj}
\ee
The expression for   $\zz_{_{\rm F}}
 (\b)$ was given in \ci{aha}  (taking the $q\to 1$ limit  and comparing to the real scalar counterpart in \rf{18}   one checks that it describes the right number of degrees of freedom).
$\zz_{\half}(\beta)$   in \rf{maj}  has  also  the  interpretation of the  conformal  group
  character for a complex  fermion  representation  \cite{dolan}.

The corresponding Casimir energy in the  Majorana or Weyl   case  is (cf. \rf{04},\rf{a1},\rf{zet})\foot{The expression
\rf{a1f}   agrees,  of course,   with   the spectrum
of the Dirac operator $-\nabla^2 +  {1\ov 4} R $   on $S^{d-1} $: the eigenvalues
are $\l_n = [ n + \ha ({d-1) }\big]^2= \om_n^2 $\  ($n=0,1,2,...$)  and their    degeneracy  is  twice
$\dd_n =  2^{[{d\ov 2}]}  { n + d -2 \choose d -2} $ (there are two $(n\pm \ha,\ha,0,...)$  representations).
For Majorana  fermions these are  to be  counted  with normalization $\ha$   relative to  a real scalar
contribution, leading
to $E_c = -\ha \sum_{n=0}^\infty   \dd_n \om_n$ equivalent to \rf{a1f}.
}
\be  \la{a1f}
E_c= - \ha \sum_n { \dd_n\, \om _n}  =- \ha \z_E (-1) \ , \ \ \ \ \ \
 \z_E (z) =     \sum^\infty _{n=0}  2^{[{d\ov 2}]} \ \te { { ( n + d-2)! \ov (d-2)! \ n!} \ \big[ n + \ha ({d-1) }\big]^{-z}} \ .
  \ee
As in the scalar case \rf{1va}, this   is  zero for odd $d$ and  non-zero for even $d$; one finds,
for instance,  $E_c=-\frac{1}{24},\frac{17}{960},-\frac{367}{48384}$ in $d=2,4,6$ respectively.\foot{This  agrees, e.g.,    with  the standard  values of the Majorana fermion Casimir energy in $d=2$  (i.e. $-{c \ov 12}, \ c=\ha$)
  and also with the value of the  Dirac fermion Casimir  energy in $d=4$:
$2 \times \frac{17}{960}$  \ci{coc}.}

\renewcommand{\theequation}{4.\arabic{equation}}
 \setcounter{equation}{0}

\section{
Free  CFT partition functions  on $S^1 \times S^{d-1}$ with  singlet constraints}
\label{singlets}

In the context of AdS/CFT duality \ci{Klebanov:2002ja}
we are  interested in conformal scalar partition   function  with an extra  singlet constraint.
As found in \ci{sy,jev},  in the case of  $N$ complex scalars  transforming in the fundamental representation of $U(N)$,
taking large $N$   limit and imposing the singlet constraint
 one effectively
 gets  instead  of \rf{18}  the {\it square}  of the one-particle partition function
\be
\zz_{U(N)} (\b) = \big[ \zz_0 (\b)\big]^2 =   {q^{ d-2 } ( 1 + q)^2   \ov ( 1 - q)^{2(d-1)} } \ . \la{19}  \ee
Below we   shall  first review the derivation of this result in \ci{sy} (which was based on \ci{Sundborg:1999ue,aha,sh})
streamlining the argument  and extending it to any dimension $d\geq 3 $. We shall then
   generalize it to the real scalar
 $O(N)$ case as this   will allow us to  compare to the  minimal HS theory  free energy in  thermal AdS$_{d+1}$.
 In the real case we shall find that\foot{The $d=3$ case of this expression was  found in \ci{jev} using
  a collective field theory approach to vectorial duality.}
\be \la{rea}
  \zz_{O(N)} (\b) = \ha   \big[ \zz_0 (\b)\big]^2   +  \ha  \zz _0 ( 2\b)=
   \ha   { q^{d-2} ( 1 + q)^2 \ov  (1-q)^{2d-2} }   + \ha  { q^{d-2} ( 1 + q^2) \ov  (1-q^2)^{d-1} } \ .
   \ee
Similarly,  in the  complex and real fermion cases we shall find
\be \la{fee}
\zz^{\rm ferm}_{U(N)} (\b) = \big[ \zz_\half (\b)\big]^2 \ , \ \ \ \ \ \ \ \ \
\zz^{\rm ferm}_{O(N)} (\b) = \ha   \big[ \zz_\half (\b)\big]^2   -  \ha  \zz _\half  ( 2\b)
\ , \ee
where $ \zz_\half (\b)$  is given in \rf{maj}.

One may also consider free theories with several fundamental flavors. For instance, we can start with $NN_f$ free complex scalars $\phi^{ia}$, $i=1,\ldots,N$, $a=1,\ldots,N_f$, and impose the $U(N)$ singlet constraint, and similarly for fermionic theories. Of course, one can also consider $NN_f$ real fields with the $O(N)$ singlet constraint. For such theories, the Casimir term is simply $F_c = N N_f \beta E_c$, where $E_c$ is the Casimir energy of a single scalar or fermion. On the other hand, the one particle partition functions that contribute to the non-trivial part of the free energy now take the form (assuming $N_f$ is fixed in the large $N$ limit)
\be \la{Nf-scalars}
\zz_{U(N)}^{N_f}(\b) = N_f^2 \big[ \zz_0 (\b)\big]^2\,,\qquad
  \zz_{O(N)}^{N_f}(\b) = \frac{N_f^2}{2}   \big[ \zz_0 (\b)\big]^2   +
\frac{N_f}{2} \zz _0 ( 2\b)
\ee
for the scalar theories, and
\be \la{Nf-fermions}
\zz_{U(N)}^{N_f-{\rm ferm}}(\b) = N_f^2 \big[ \zz_\half (\b)\big]^2\,,\qquad
  \zz_{O(N)}^{N_f-{\rm ferm}}(\b) = \frac{N_f^2}{2}   \big[ \zz_\half (\b)\big]^2   -
\frac{N_f}{2} \zz _\half ( 2\b)\,,
\ee
for the fermion ones. These theories are dual to versions of Vasiliev higher spin theory with $U(N_f)$ or $O(N_f)$ bulk gauge symmetry \cite{Vasiliev:1999ba, Chang:2012kt}. In Section \ref{HS-Nf} we will show how the  above  thermal partition functions, as well as the Casimir energies, are reproduced by the sums over higher spin fields in AdS$_{d+1}$.

 \subsection{Complex scalar case}

 Starting  with  $N$  complex  scalars   of   fundamental representation of $U(N)$,  to ensure the  singlet condition
 one may couple  them  to a  constant  flat connection $A_0 = U^{-1} \del_0 U$  with $U \in U(N)$
 and integrate over $U$.
 The resulting scalar  operator will   have eigenvalues  as in \rf{111}   but with $k$   shifted by  phases $\a_i$
 of the eigenvalues $e^{i \a_i}$ ($i=1, ..., N$)
 of the holonomy matrix, i.e.  $\l_{k,n}= ({2 \pi k + \a_i \ov \b} )^2  +   \omega^2_n$.
 The   resulting scalar determinant  is then  to be integrated  over $\a_i$   with  the standard  $U(N)$ invariant measure
  given by the Van der Monde   determinant, $[dU]= \prod^N_{k=1} d \a_k\ \prod^N_{ i \not=j=1}  |e^{i\a_i} - e^{i\a_j}|   $ (see, e.g., \ci{meh}).
As a result, the singlet partition function $\hat Z$   is the following modification of  $Z$   in \rf{11},\rf{14} (for  $2N$ real scalars) \ci{sh,sy}
\be
&& \hat Z = e^{ - \hat F}  \ , \ \ \ \    \hat F = \hat F_c  + \hat F_\b   \ , \ \ \  \ \ \ \ \ \
   e^{ - \hat F_\b}  = \int  \prod^N_{k=1}  d\a_k \ e^{ - \tilde F_\b (\a_1,...,\a_N)} \ ,                  \la{20}  \\
&&  \la{201}
\tilde F_\b = -\ha  \sum^N_{i\not= j=1} \ln \sin^2{\te  { \a_i - \a_j \ov  2}}    +   2 \sum_{i=1}^N  f_\b (\a_i) \ , \\
&&f_\b (\a) =   \sum_{m=1}^\infty   c_m(\b)   \cos (m\a) \ , \ \ \ \ \ \ \ \ \ \ \ \ \    c_m (\b) =  - { 1 \ov m} \zz_0 (m \b)\ .   \la{21}
\ee
Here $\zz_0 (\b)$ is the same    one-particle   partition function as in \rf{18}  (``one-letter'' partition function of \ci{aha})
so that in the formal limit of $\a_i\to 0$  we get  $\tilde F_\b  $ reducing 
to  $ 2N$  times   free energy of a real scalar  $F_\b$  in \rf{17}.
The ``trivial'' (not sensitive to $\a_i$ averaging)  Casimir  part
$ \hat F_c $   
is  the same as in \rf{a1},\rf{15}  up to the $2N$ factor
\be
 \hat F_c = 2 N \b  E_c  \ . \la{ca} \ee
In the large $N$ limit $ \hat F_c$   scaling  as $N$   should match  the   contribution of the
 classical  higher spin  action  in the AdS$_{d+1}$ bulk.
At the same time, the nontrivial  part of $\hat F_\b $  which  will   happen   to scale as $N^0$  due to extra averaging over $\a_k$
 \ci{sy}
 should thus  be matched  with  the  1-loop
 partition function of HS   theory   in thermal AdS$_{d+1}$.

Considering the large $N$ limit one introduces  as usual  the eigenvalue density $\r(\a), \ \a\in (-\pi, \pi)$ and  replaces  the integral
over $\a_i$ by the path integral over  the perdiodic  field  $\r(\a)$ defined  on a  unit circle  with  the action
\be\la{22a}
&&  \tilde F_\b (\r) =    N^2 \int d\a d \a'\   K(\a- \a') \r(\a) \r(\a')    +  2 N  \int d \a \, \r(\a) f_\b (\a) \ , \\
&& \la{22b}
K(\a) = -   \ha   \ln  \big(2 - 2 \cos\a\big) 
\ , \ \ \ \ \ \ \ \ \ \ \ \ \ \ \
f_\b (\a)  =   \sum_{m=1}^\infty  c_m(\b)     \cos (m\a) 
\ . \ee
Note that the fact the second term in (\ref{22b}) scales as $N$ is because the matter is in the fundamental representation. Then, in the large $N$
limit and as long as the temperature is parametrically smaller than a power of $N$ \cite{sy}, the saddle point solution for the eigenvalue density takes the form $\rho(\alpha)=\frac{1}{2\pi}+\frac{1}{N}\tilde\rho(\alpha)$, where $\tilde\rho(\alpha)$ does not contain a constant part.\foot{The constant mode of $\rho$
ensures that  it satisfies the normalization condition $\int_{-\pi}^{\pi} d\alpha \rho(\alpha)=1$.}
An important point is that the constant part of $\r(\a)$ does not couple to the source $f_\b$ (which does not contain a zero mode term)
so that it can be effectively projected out without changing the non-trivial $\b$-dependent part of the result.
This allows, in particular, to ignore the constant part   of $K$.\foot{Note that up  to  the  factor of $1\ov \pi$  the  kernel $K$  is the
same as  the  restriction of the   Neumann function on a unit disc to its boundary.}
Then, doing the formal gaussian path integral over periodic non-constant  $\r(\a)$
gives an $N$-independent result for $\hat  F_\b$ in \rf{20}
\be
\hat  F_\b    = -   \int d\a d \a'\   K^{-1} (\a- \a') f_\b (\a) f_\b (\a')   \  . \la{23}
\ee
The kernel $K$  and its inverse $K^{-1}$
 have  simple  Fourier expansions\foot{We used
the identity
$ \ln (1   +   b^2 - 2 b \cos \a) =  -2 \sum_{m=1}^\infty { b^m \ov m } \cos ( m \a ) $. Note also
 that the  delta-function  on  non-constant  functions on a circle
is $\bar \delta (\a) = { 1 \ov \pi}  \sum_{m=1}^\infty  \cos ( m \a )$   with
$\int d \a\, \cos (m \a) \ \cos (n \a ) = \pi \delta_{mn}$.}
\be\la{kk}
K(\a) =   \sum_{m=1}^\infty { 1 \ov m}   \cos ( m \a ) \ , \ \ \ \ \ \   \ \ \ \ \ \
K^{-1} (\a) =  {1 \ov  \pi^2}  \sum_{m=1}^\infty { m}   \cos ( m \a ) \ .
\ee
We conclude that \rf{23}  is given by
\be\la{c2}
  \hat  F_\b   = -    \sum_{m=1}^\infty  m  \big[c_m(\b)\big]^2 \ , \ee
    where $c_m(\b)$ was defined in \rf{21},  or   explicitly
 \ci{sy}
\be
\hat  F_\b    =  - \sum_{m=1}^\infty   { 1 \ov m}   \zz_{U(N)} (m \b) \ , \ \ \ \ \ \   \ \ \ \ \ \    \zz_{U(N)} (\b) =   \big[ \zz_0 (\b)\big]^2     \  .  \la{24}
\ee
Compared to  the ``unprojected''  free energy of a single real scalar  in \rf{17}   here one gets
   the second power of  the  one-particle  partition function
$\zz_0$ factor in \rf{18},\rf{19}. As we have seen above, this squaring of $\zz_0$
has its origin in the gaussian  averaging  over the  density of  the large $N$  distribution of   the   eigenvalues of the
holonomy matrix. In Section \ref{therm-HS}, we will see that this result precisely matches the one-loop partition function of the higher spin theory in thermal AdS. Note that
an important difference compared to the Yang-Mills case \cite{aha} is that in these vectorial models there is no phase transition at temperatures $T\sim 1$ (i.e. temperatures of order of the AdS scale). A phase transition only occurs at much higher (Planck scale) temperatures $T \sim N^{\frac{1}{d-1}}$ \cite{sy}, where the calculation above breaks down (see the discussion below eq.~(\ref{22b})).

\def \N {{\rm N}} \def \K {{\rm K}} \def \k {{\rm k}}

 \subsection{Real scalar case}
\label{real-scalars}

Let us now  repeat  the   above  discussion in
 the  case of $N$ real scalars  transforming as a fundamental representation of $O(N)$.
Since we are interested only in the large $N$ limit we may  choose $N$ to be even, $N=2\N$.\footnote{We expect that the $1/N$ expansion for $O(N)$ should not be 
sensitive to whether $N$ is even or odd. See for instance \cite{Sinha:2000ap} for an explicit example of the large $N$ expansion of $SO(N)$ Chern-Simons theory.}
An orthogonal $N \times N$   matrix can   be put into a  canonical   form  with $\N$   diagonal $2\times 2$    blocks
$ {\cos \a \ \ -\sin \a} \choose  { \sin \a \ \ \cos \a} $   which can   be further  diagonalized to
$ {e^{i\a} \ \ 0} \choose  {\  0 \ \  e^{-i \a}} $.  This   may  be formally viewed  as a special  $U(N)$ case
where $N$   eigenvalues $\a_i$ are chosen as $(\a_1, -\a_1, \a_2, -\a_2, ..., \a_\N, -\a_\N)$.
Then the analog  of $\tilde F_\b$ in  \rf{201}   becomes  (here we  use $r,s= 1, 2, ..., \N$  to label  $\a_i$ from different 2-planes)
\be \la{r1}
\tilde F_\b = - \ha  \sum^\N_{r\not=s=1} \ln   \sin^2{\te  { \a_r - \a_s\ov  2}}   - \ha  \sum^\N_{r,s=1}    \sin^2{\te  { \a_r + \a_s\ov  2}}
 +  \ha \sum^\N_{r =1} \ln \sin^2{  { \a_r }}      +   2 \sum_{r=1}^\N  f_\b (\a_r) \ ,
\ee
where  $f_\b (\a_r)$ is given by the same  expression as in \rf{21} (that for $\a_r=0$
the last term  becomes $N f_\b(0)$   or $N$ times $F_\b$ in \rf{17}  as  it should be for $N$ real scalars).
The Casimir energy term is the same as in \rf{ca} with $N \to \N$.
In the large $N$ limit \rf{r1} is then replaced by
\be
\la{22r}
 && \tilde F_\b (\r) =    \N^2 \int d\a d \a'\   \K(\a,  \a') \r(\a) \r(\a')    +  2 \N \int d \a \,  \r(\a)  \k(\a)     +  2 \N  \int d \a \, \r(\a) f_\b (\a) \ ,  \ \ \ \    \\
 \la{223r}
&& \K(\a,\a') =  -   \ha   \ln \Big(  \big[ 2  - 2 \cos (\a-\a')  \big] \big[ 2  - 2 \cos (\a+\a')  \big]\Big) 
= 2 \sum_{m=1}^\infty { 1 \ov m}   \cos (m \a) \,   \cos (m\a')
\  , \ \ \ \  \ \no  \ \\
&& \ \ \ \ \qquad    \k(\a)=   \fo   \ln (2 - 2\cos 2 \a) = -\sum_{m=1}^\infty { 1 \ov 2m}   \cos (2 m \a ) \la{24r}
\ . \ee
Since $f_\b (\a)$ in \rf{21}  contains only $\cos m \a$ modes in its expansion,   the constant
and $\sin m \a$ modes of a generic periodic function $\r(\a)= a_0 +  \sum_{m=1}^\infty ( a_m \cos m \a + b_n \sin m \a) $
do not couple to the $\b$-dependent source, i.e.     we may restrict  the integration to even non-constant functions $\r(\a)
=   \sum_{m=1}^\infty  a_m \cos m \a$ (this allows,  in particular,  to ignore constant terms
in $\K$ and $K$).
 Then the  gaussian integration  gives   again an order $N^0$ term (cf. \rf{23})
\be \la{g1}
&&\hat  F_\b    = -   \int d\a d \a'\   \K^{-1} (\a, \a')\,  j(\a)\,  j (\a')  \ ,\la{32r}   \\
&&\K^{-1} (\a, \a') = { 1 \ov 2 \pi^2}  \sum_{m=1}^\infty {  m}   \cos ( m \a ) \cos (m \a')   \ ,  \la{234r}\\
 &&j =  f_\b (\a) +  \k(\a)= \sum^\infty _{m=1,3,5, ...} c_m   \cos (m \a)
+ \sum^\infty _{m=2,4,6, ...}  c'_m   \cos (m \a)  \ ,  \la{277r}  \\
&& c_m= -{1\ov m} \zz_0( m \b) \ , \ \ \ \ \    \ \ \ \ \  c'_m = c_m - { 1 \ov m }  =
-{1\ov m}  \big[  \zz_0( m \b)    + 1   \big] \ ,
 \la{27r} \ee
where we used    \rf{21}.
As a result,  we find in the real scalar  case
(omitting $\b$-independent constant)
\be \la{g11}
&&\hat  F_\b{}  = -  \sum_{m=1}^\infty { 1 \ov m} \zz_{O(N)} (m\b)  \ , \ \ \ \ \ \ \ \ \ \
 \zz_{O(N)} (\b) = \ha   \big[ \zz_0 (\b)\big]^2  +  \ha  \zz _0 ( 2\b) \ . \ee
 The second term in $ \zz_{O(N)}(\b)$ originates from the  cross-term between the  source $\k(\a)$
 coming from the measure  and  the  $\b$-dependent source   $f_\b (\a)$.
 The explicit form $\zz_{O(N)}(\b)$ was already  given in \rf{rea}.

If the case of $N_f$ free complex or real scalar flavors in the fundamental of $U(N)$/$O(N)$, with $N_f$ fixed in the large $N$ limit, the only modification is that the last term in (\ref{22a}) and (\ref{22r}) acquires an extra factor of $N_f$. Then, the same calculation as described above readily leads to the results in (\ref{Nf-scalars}) for complex or real scalars.

\def \half  { {1\ov 2} }
 \subsection{Fermionic theories}
\label{singlet-fermions}

In the $U(N)$   invariant  case of $N$   Dirac fermions  the singlet constraint  is  again implemented by
averaging the  Dirac operator determinant over  the $U(N)$  holonomy  eigenvalues \ci{sh,sy}.
One difference compared to the scalar case  in \rf{20} is that now we will have the Casimir  part of free energy \rf{ca}
 replaced   by its fermion  analog; also,
 the non-trivial $\b$-dependent  term $f_\b$  in \rf{21}  will be replaced by
\be
f_\b (\a) =   \sum_{m=1}^\infty   c_m(\b)   \cos (m\a) \ , \ \ \ \ \ \ \ \ \ \ \ \ \
   c_m (\b) =  - { (-1)^{m+1} \ov m} \zz_\half  (m \b)\ ,    \la{21f}
\ee
where $\zz_\half $ was defined in \rf{maj}, i.e. one is to replace the real scalar  one-particle partition
function $\zz_0$ in \rf{21}   by  real  or Weyl  fermion partition function $\zz_\half $ (and add an extra $(-1)^{m+1}$ factor).\foot{The normalization   can be  checked  by considering the $U(1)$ case  when the free energy of  a  single complex scalar
or $F_\b =- 2 \sum_{m=1}^\infty  { 1 \ov m } \zz_0 ( m \b)$  (cf. \rf{17}) should be replaced  by   free energy of a single Dirac
fermion, or    $F_\b =- 2 \sum_{m=1}^\infty  { (-1)^{m+1}  \ov m } \zz_\half  ( m \b)$  (cf. \rf{17f},\rf{maj}).
An extra factor of 2 in eq. 24  of \ci{sy}  appears to be a misprint.}

 The rest of the argument  is unchanged, so we  again end up with \rf{c2}, now with $c_m$ given in \rf{21f},  i.e.
 \be  
\hat  F_\b    =  - \sum_{m=1}^\infty   { 1 \ov m}   \zz^{\rm ferm}_{{U(N)}} (m \b) \ , \ \ \ \ \ \   \ \ \ \ \ \
   \zz^{\rm ferm}_{{U(N)}}(\b) =   \big[ \zz_\half (\b)\big]^2     \  .  \la{24f}
\ee
In the $O(N)$  singlet sector of $N$ Majorana fermions, the starting point is  \rf{r1}
with the same function $f_\b$ as in \rf{21f}.
Doing the obvious replacement in \rf{27r}, i.e.
$c_m= {(-1)^m \ov m} \zz_\half ( m \b)  , \ \  c'_m = c_m - { 1 \ov m }  $
(so that  $c'_{2m} = { 1 \ov 2m } \big[  \zz_\half ( 2 m \b) - 1 \big]$)
  we finish with the following analog of  \rf{g11}
\be \la{gf}
\hat  F_\b{}  =  -  \sum_{m=1}^\infty { 1 \ov m} \zz^{\rm ferm}_{O(N)} (m\b)  \ , \ \ \ \ \ \ \ \ \ \
 \zz^{\rm ferm}_{O(N)} (\b) = \ha   \big[ \zz_\half  (\b)\big]^2  -   \ha  \zz _\half  ( 2\b) \ .  \ee
We shall see  in Section \ref{free-fermions}  how    the  partition functions in \rf{24f} and \rf{gf}  are reproduced
on the  dual  AdS  higher spin  theory  side.

While the above calculation was presented in the case of a single fundamental fermion, it is straightforward to generalize it to the case of $N_f$ flavors ($N_f\ll N$). One simply includes an extra factor of $N_f$ in the free fermion one-particle partition function (\ref{21f}), and performing the Gaussian integral over the eigenvalue density immediately leads to the results quoted in (\ref{Nf-fermions}).

\renewcommand{\theequation}{5.\arabic{equation}}
 \setcounter{equation}{0}

\section{Higher spin partition function in AdS$_{d+1}$  with $S^1 \times S^{d-1}$  boundary}
\label{therm-HS}

Our aim  in this   section  will be
  to compare the singlet sector scalar  CFT  free energies, calculated above,  with their  counterparts
for  higher spin   theories in  thermal  AdS$_{d+1}$.  The   HS  free energy  can  be found by summing
the individual massless  spin $s$ contributions:\foot{Here we consider free massless totally symmetric  higher spins
with Lagrangian originally found by Fronsdal in AdS$_4$ \ci{Fronsdal:1978vb}.
An extension of Fronsdal work to higher dimensions was carried out  in \ci{Metsaev:1994ys,Metsaev:1997nj}.}
\be   F = \sum_s F^{(s)}  \ , \ \ \ \ \     F^{(s)}  = -\ln Z_s \ , \ \ \ \ \ \
 Z_{s}  =
 \left({\det\, \big[-\nabla^2  + (s-1) ( s + d-2) \big]_{s-1\, \pe} \ov
 \det\, \big[-\nabla^2  -  s + (s-2) ( s + d-2) \big]_{s\pe}}\right)^{1/2} ,\  \ \    \la{1}
 \ee
 where  the  operators are defined on symmetric traceless  transverse  tensors \ci{gl,Gaberdiel:2010ar}.\foot{We set the AdS scale to 1.  The energies (or scaling dimensions) of the corresponding representations are
$\Delta=e_0= s + d-2$    for the  physical field  \ci{Metsaev:1994ys}  and $\Delta=e_0 = s + d-1 $ for the ghost one.}

On general grounds, for a quantum   field  in   AdS$_{d+1}$  with boundary $S^1_\b \times S^{d-1}$
 one expects  the 1-loop  free energy  to have the following structure  (cf. \rf{14})
 \be
 F = F_0  + F_\b \ , \ \ \ \ \ \ \ \ \ \ \ \  F_0 = \b \bar F_0  =F_\infty  +  F_c \ , \ \ \ \ \ \ \ \ \ \ \
 F_\infty =  a_{d+1}   \ln \Lambda\ ,
  \la{51}\ee
  where  $F_0$  is linear in $\b$  (i.e. proportional to the volume)  
  while  the  part $F_\b$  with  non-trivial  $\b$ dependence
   is finite and vanishes in the zero-temperature limit $\b \to \infty$.
    We have split $F_0$ into a possible  UV  logarithmically divergent part, and a finite part $F_c$
    (power divergences are  assumed to  be regularized away).

The    coefficient  $ a_{d+1}$ of the UV divergent term  vanishes automatically if $d+1$ is odd.
 For even $d+1$   it is  given by an integral of the corresponding  local Seeley coefficient,   which,  in
 the case of  AdS$_{d+1}$,  is  proportional to the   product of the  volume  factor  and  $\ha ( d+1)$   power of  the
  constant    curvature.  Since this curvature factor  is  the same for any $\b$ or regardless of the topology of the boundary,
  the dependence   of $a^{(s)}_{d+1}$   on the  spin $s$     should be the same    as
  that found  in \ci{gk,gks}  for the case of  Euclidean  AdS$_{d+1}$, i.e. the hyperboloid $H^{d+1}$,  whose boundary is
   $S^d$.  In particular, it was  shown in  \ci{gk,gks}   (for  various   values of $d$)
   that the total  anomaly coefficient
   $\sum_s  a^{(s)}_{d+1}$  {\it vanishes}  after   summing
   over spins  (assuming  the zeta-function regularization of the sum),
  so that  there are no  logarithmic UV divergences in  the  standard or minimal    HS theory.
Thus  in what follows we shall set $F_\infty=0$.

\def \Vol {{\rm Vol}}

The problem of computing  the $\b$-dependent part  $F_\b$   of the one-loop   free energy  in thermal AdS$_{d+1}$
can be approached   from  the Hamiltonian  point of view \ci{ad,gpp},
  using group-theoretic   considerations to determine the  energy
spectrum \ci{ad} of a  spin $s$  field  in  global AdS$_{d+1}$  with  reflective  boundary
 conditions \ci{Avis:1977yn,Breitenlohner:1982jf}.
An equivalent result  for $F_\b$  is  found
in the path integral approach  by starting with the heat kernel  for the  hyperboloid $H^{d+1}$
 \ci{Camporesi:1991nw,Camporesi:1992wn,Camporesi:1993mz}
 and using the method of images to find  its  counterpart for thermal  AdS$_{d+1}$
 viewed as a  quotient  $H^{d+1}/Z$
 (see  \ci{Giombi:2008vd,David:2009xg} for the AdS$_3$ case
  and  \ci{ggl,gl}  for  the general case).\foot{Ref. \ci{ggl}  used, in fact,
  an analytic continuation of a  heat kernel  of a quotient of a sphere   $S^{d+1}$.}
 In this heat kernel approach $F_0$ in \rf{51} is the ``zero-mode''  part of the sum over the images, and
 it is  thus natural to identify it with  $\hat F_0= {\Vol(H^{d+1}/Z)\ov \Vol(H^{d+1})}  F(H^{d+1})$, where
 $F(H^{d+1})$ is the free energy on $H^{d+1}$. This  $\hat F_0$
   requires  a proper definition or regularization (cf.  \ci{Giombi:2008vd})
   and  was not studied  in \ci{ggl}.\foot{For a  related recent discussion  of    $\hat F_0$  in the case of
   a  massive scalar   in AdS$_2$  and AdS$_4$    see \ci{ke}.}

At the same time,
the expected  correspondence with  the free energy \rf{14} of the boundary CFT  in $S^1 \times S^{d-1}$
suggests that   the finite part of $F_0$  should   be  closely related to the  vacuum or Casimir energy
of the corresponding fields  in AdS$_{d+1}$.  As this relation  appears to be
 obscure in the  $H^{d+1}/Z$   construction
let us   discuss  an alternative approach to justify it.

 Let us recall that starting with the Euclidean  AdS$_{d+1}$, realized as a hyperboloid
   $x^2_{d+1} - x_0^2 -  x_i x_i = 1$  in $R^{1 , d+1}$,
  one may   choose different sets of coordinates  (see, e.g., \ci{Gibbons:2011sg}).
  For example,  one  may
  set $x_{d+1} = \cosh \xi , \   x_m = ( x_0, x_i) =  \sinh \xi\, n_m$
  where $n_m n_m=1$   getting the  $H^{d+1}$  metric
  $ds^2_\xi  = d \xi^2 + \sinh^2 \xi\ d\Omega_d$   with $S^d$  as its  boundary. One may  also choose
  the coordinates as   $x_{d+1} = \cosh \r\, \cosh t , \   x_0=  \cosh \r\, \sinh t , \ x_i=   \sinh \r\,  n_i, \ \  n_i n_i=1$
   obtaining  the Euclidean
    continuation of the   global AdS$_{d+1}$ metric, i.e.
      $ds^2_t = \cosh^2 \r \, dt^2  +   d \r^2  + \sinh^2 \r \, d \Omega_{d-1}$.  
   Compactifying $t$ on a circle of length $\b$ gives thermal AdS$_{d+1}$
     metric $ds^2_{t,\b}$   with  $S^1_\b  \times S^{d-1}$  as the
     boundary. A  direct computation of  the {full}  expression for the  determinant of a scalar Laplacian
       in the case of AdS$_{d+1}$  with $ds^2_t $ or $ds^2_{t,\b}$   metric did   not  seem to
     appear    in the literature.
     In particular, it is not obvious    how   a  finite part  $F_c$   of  the  $\b \to \infty$ limit  of
     $F$   computed   for $ds^2_{t,\b}$   will
     match the expression found   in  \ci{Camporesi:1991nw,Camporesi:1993mz}  in the case of the hyperboloid $ds^2_\xi$.

    The    AdS$_{d+1} $ metric   $ds^2_t$  may  be written also as
  $ ds^2_t = {1\ov \cos^2 \theta} ( dt^2 +   d \theta^2 + \sin^2 \theta\,  d\Omega_{d-1})$,
  where $\theta \in [0, {\pi\ov2})$  with the boundary $R \times S^{d-1}$ at $\theta={\pi \ov 2} $. Thus, it is
    conformal to   a  half  of  the Einstein universe  $R \times S^{d}$.
        Indeed,  there is a close correspondence   \ci{Dowker:1983nt}  between  the  spectrum of  the
  energy  operator in AdS$_{d+1}$ and the
 spectrum of the Laplacian in the  half of the Einstein universe with
 reflective boundary conditions  (Dirichlet or Neumann)  at  the equator of $S^{d}$
 \ci{Avis:1977yn,Breitenlohner:1982jf}.
 In particular, the Casimir energies
  in  AdS$_4$ (as defined in \ci{ad})   and  in  $R\times S^3$  are  the same  up to a  factor  \ci{Dowker:1983nt}.
  This implies  also  that  this  energy spectrum  determines
    the non-trivial part  $F_\b$ of the free energy,\foot{
    Note that for  conformally invariant fields $F_\b$   is always the same in conformally related static spaces \ci{Dowker:1978md}.}
     in agreement with  its alternative derivations
     in \ci{Allen:1986ty} (from direct  evaluation of the scalar stress tensor in AdS)
  and in  \ci{Giombi:2008vd,Gaberdiel:2010ar,ggl}  (from  the $H^{d+1}/Z$ construction of the heat kernel).\foot{It should be
   noted that while  one  may  expect  the vacuum energy to scale as  volume of global AdS   space
(which should factorize as AdS is a homogeneous space) this is
actually in contradiction with reflective energy-conserving   boundry conditions  (appropriate for finite temperature set up)
which lead  to  discrete  spectrum  of the Laplacian (see  \ci{Sakai:1984fg} for a discussion in AdS$_2$ case).
We expect that under an appropriate regularization,  the large  $\b$ limit of  the stress   energy computation in
 \ci{Allen:1986ty}  should   reproduce  the ``non-extensive''
 expression for the total  AdS vacuum energy as a  sum over global energy eigenvaluaes \rf{02}
 found in the Hamiltonian approach \ci{ad,gpp}. One possibility
 may be to  do the integration over the radial AdS direction for finite $\b$ and then take the limit $\b\to \infty$ in the result.
 }

It remains then to  understand the relation between  the $F_0$ part of   free energy in AdS$_{d+1} $
and  in  the  conformally-related Einstein universe.
Given an ultrastatic space-time   $ R \times M^{d}$  with  a Euclidean  metric
  $ds^2 =  dt^2 +   \tilde g_{ij}(x) dx^i dx^j$,
 one can readily show that, up to a standard  logarithmically divergent term proportional to $\zeta(0)$   \ci{Allen:1986qi,my},
  the corresponding free energy or $-\ln Z$
     is given by the Casimir energy term $F_c=\b E_c$. Here
 $\b\to \infty $ is the length of  the time interval  and $E_c= \frac{1}{2}\sum_n \dd_n \om_n$,
 with $\om^2_n$  being  the eigenvalues of  the Laplacian restricted to $M^d$ (cf. Section \ref{back}).
For a  conformally-related   static space-time    $ds^2 =  g_{00} (x)  dt^2 +   g_{ij}(x) dx^i dx^j $
 the full  expression for the free energy
  will  contain,  in addition to $F_c$,
    extra  $g_{00}$-dependent  local terms
 reflecting     the required conformal rescaling by $g_{00}(x)$    \ci{Dowker:1978md,Dowker:1989gp}
 (see also, e.g.,   \ci{Kirsten:1991mi,Camporesi:1992wn}).

 These  extra terms are linear in $\b$, i.e. not changing the $F_\b$ part in the finite temperature case.
These  terms are similar   \ci{Dowker:1989gw}
 to the ``integrated conformal anomaly''   terms found  for  conformally invariant matter fields. They  should
  be closely related to the $\zeta(0)$-type terms  that  contribute to $F_\infty$ part of $F$ in \rf{51}
  and should thus  vanish like   $\sum_s  a^{(s)}_{d+1}$   after  one  sums over  the spins.
  For that reason in what follows we shall assume
  that  the $\b \to \infty$ limit of the UV  finite part    of the  free energy \rf{51}  in thermal AdS$_{d+1}$
 has indeed   the interpretation of the Casimir energy term, i.e.  $F_c=\b E_c$.
 As we shall see below, this is   fully consistent with the  AdS/CFT correspondence.

 \subsection{Temperature-dependent part of the free energy}
\label{scalar-Fbeta}

Let us  first discuss the temperature dependent part,  $F_\b$, of the  free energy of higher spin theories
and then turn to the Casimir part in the  next subsection.

 The  expression    for   $F_\b$  of totally symmetric massless  spin $s$ field is
 \ci{ggl,gl}
\be
&&  F^{(s)}_\b = -\sum_{m=1}^\infty { 1 \ov m}   \zz_s (m \b) \ , \la{2} \\
&& \zz_s ( \b)  = {q^{s + d-2} \over (1 - q)^{d} } \big(  \dd_s - \dd_{s-1} q\big)  \ , \ \ \ \ \ \ \ \  \ \ \ \ \   q\equiv e^{-\b}  \la{3}\\
&& \dd_s = ( 2 s + d-2) { (s+ d -3)! \ov (d-2)!\,  s!}  \ . \la{4}
\ee
 Here $\dd_s$ is the number of symmetric traceless rank $s$ tensors in $d$ dimensions or dimension of $(s,0,0,...)$  representation of
 the ``little'' group $SO(d)$
(i.e. $\dd_s{\big|_{d=3}} =2s+1    , \ \ \dd_s{\big|_{d=4}} =(s+1)^2   \ , $ etc.).
Note that $\dd_s$  is exactly the same as the degeneracy  of eigenvalues of the scalar Laplacian on $S^{d-1}$
  if we replace the angular momentum quantum number $n$ in  $\dd_n$  in \rf{c0c}  by the spin  $s$.

 $\zz_s ( \b) $  in \rf{3}  thus  has  an   interpretation of  the   corresponding
  one-particle partition function (cf. \rf{03},\rf{06}). From the point of view of $d$-dimensional CFT, $\zz_s$ is the character of the representation
  of $SO(d,2)$ containing the spin-$s$ primary field of dimension $\Delta=s+d-2$ and its descendants \cite{dolan,gpp}. 
The explicit results for $d=2,3,4$ are
\be
&&d=2: \ \ \ \ \ \     \zz_{s>1 }  ( \b)  = {2q^{s} - 2 q^{s+1}\over (1 - q)^{2} }= {2 q^s\over (1-q) }\ , \ \ \ \ \ \ \ \ \  \label{dtwo} \\
&&d=3: \ \ \ \ \ \     \zz_{s>0 }  ( \b)  = {q^{s+1} \over (1 - q)^{3} } \Big[  2 s+1  -  (2s-1)  q \Big] \ , \ \ \ \ \ \ \ \ \ \label{5} \\
&&d=4: \ \ \  \ \ \    \zz_{s }  ( \b)  = {q^{s+2} \over (1 - q)^{4} } \Big[  (s+1)^2  -  s^2   q \Big] \ . \la{6}
\ee
The low spin cases in $d=2,3$ are special. For the spin $0$ primary field of general dimension $\Delta$,
\be
 \zz_{0}^{(\Delta)}= {q^\Delta \over (1-q)^d}\ ,
\ee
since no ghosts need
to be subtracted (this  ghost term vanishes automatically for $s=0$ for all $d>3$). In AdS$_3$ there are two possibilities for $s=1$. For the Maxwell action, which
was conjectured to be relevant to the $d=2$ scalar theory \cite{gks},
$\zz_{1}^{\rm Maxwell}={2q-q^2\over (1-q)^2}$; for the Chern-Simons action, which is relevant to the $d=2$ fermionic theory,
$\zz_{1}^{\rm CS}={2q-2q^2\over (1-q)^2}$.

Putting these elements together, we find the total free energy of the standard Vasiliev theory in AdS$_{d+1}$ where
each integer spin is counted once \cite{Vasiliev:2003ev}. Including the $s=0$ contribution with $\Delta=d-2$, and $s=1$ Maxwell theory for $d=2$,
we find that that in all dimensions $d\geq 2$
\def \ZZ {{\rm Z}}
\be
&&
F_\b = \sum_{s=0}^\infty  F^{(s)}_\b    = - \sum_{m=1}^\infty { 1 \ov m}   \zz (m \b)  \  , \la{7} \\
&&
 \zz (\b) = \zz_{0}^{(d-2)}+ \sum_{s=1}^\infty  \zz_s ( \b) = { q^{d-2} ( 1 + q)^2 \ov (1-q)^{2d-2} } \ . \la{8}
 \ee
Comparing to \rf{24},\rf{19},\rf{18}  we conclude   that this
  is   exactly the same as the    order $N^0$   term in the large $N$ limit of the free energy  of a complex $U(N)$   scalar
in $S^1 \times S^{d-1}$  with the singlet condition imposed. This is thus a  generalization to all dimensions  $d \geq  2$ of  the matching  found
 in \ci{sy} (and also,   in the collective field theory approach, in \ci{jev}) in the  $d=3$ case.
 This   matching is  a consistency check that the boundary and the bulk theories have the same
 spectrum of  states:   indeed,  the free spectra   determine   both  the one-loop term  in  the $\b$-dependent  bulk theory free energy
 and  also the  leading order $N^0$  term  in the boundary theory  free energy.

The  identity    (\ref{8})
 also has  an interpretation as expansion of the CFT partition function in terms of characters $\zz_s$ of the conformal group.
This expansion is completely determined by the spectrum of primary fields and by the conformal symmetry. For example, if we consider the large $N$ limit of the
$d=3$ critical $U(N)$ model, then the dimension of the singlet scalar primary operator $\bar\phi_i \phi^i$ changes from $1$ to $2$, while the dimensions of
singlet higher spin primaries
remain the same as in the free theory.\footnote{In the critical vector model, the anomalous dimension of the spin $s$ currents is of order $1/N$.} Therefore, the partition function of such a large $N$ CFT must have the form\footnote{While this is guaranteed
on general grounds, it would be nice to give a direct path integral derivation.}
\be\la{cri}
\zz_{\rm crit}(\beta)= {q^2\over (1-q)^3} + \sum_{s=1}^\infty  {q^{s+1} \over (1 - q)^{3} } \Big[  2 s+1  -  (2s-1)  q \Big]
\ .
\ee
This is equal to the one-particle partition function of the Vasiliev theory in AdS$_4$ with the $\Delta=2$ boundary condition for the bulk scalar.
Therefore, once again, the AdS/CFT agreement of the partition functions follows from the conformal symmetry and the agreement of the spectra.

 Let us note   that the  HS partition function  corresponding to  \rf{7},\rf{8}     may be   rewritten as\foot{One  is to use that
 $
 \sum_{m=1}^\infty   { 1 \ov m}   {  q^{m a} \ov (1 - q^m)^b}   = - \sum_{n=1}^\infty  {n + b -2\choose b-1}  \ln ( 1 - q^{ n + a  -1})
$
since  $(1-x)^{-b} =  \sum_{n=1}^\infty  {n + b -2 \choose b-1}  x^{n-1} $.
}
 \be
\no  F_\b=-\ln Z_\b &=&
   \sum_{n=1}^\infty  {\textstyle {n + 2d  -4\choose 2d -3}  } \ln \Big[ ( 1 - q^{ n + d  -3}) ( 1 - q^{ n + d  -2})^2  ( 1 - q^{ n + d  -1})  \Big]\\
&=&    \ln  \Big[( 1 - q^{  d-2})   ( 1 - q^{  d  -1})^{2d} \Big]
  + \sum_{n=1}^\infty C_n   \ln   ( 1 - q^{ n + d  -1})  \ ,  \la{10}   \\
  C_n &=&  {\textstyle {n + 2d  -4\choose 2d -3}  } + 2  {\textstyle {n + 2d  -3\choose 2d -3}  } +  {\textstyle {n + 2d  -2\choose 2d -3}  }
  ={\te { ( n + 2d -2)! \ov  ( 2d-3)! (n+1)!}  \big[ 4 n ( n +   2d -2)  +  2d ( 2d-3) \big]} \ .
   \no     \la{10a} \
 \ee
 This   generalizes to any $d$ the  expressions given for $d=3,4,6$ in \ci{gl}.

In  the  {\it  minimal}  Vasiliev theory in AdS$_{d+1}$, which should be dual to the $O(N)$ singlet sector of the
$d$-dimensional real scalar theory,
  one is to sum over all {\it even}  spins only. Then instead of \rf{7},\rf{8} one   finds from \rf{3},\rf{4}
 \be
&&F_\b{}_{\rm\, min} = \sum_{s=0,2,4,..}^\infty  F^{(s)}_\b    = - \sum_{m=1}^\infty { 1 \ov m}   \zz_{\rm min} (m \b)  \  , \la{7e} \\
&&  \zz_{\rm min} (\b) = \zz_{0}^{(d-2)} + \sum_{s=2,4,...}^\infty  \zz_s ( \b)
  =
\ha   { q^{d-2} ( 1 + q)^2 \ov  (1-q)^{2d-2} }   + \ha  { q^{d-2} ( 1 + q^2) \ov  (1-q^2)^{d-1} } \ .
   \la{777} \ee
   This nicely matches the order $N^0$  term  \rf{g11},\rf{rea}   in the free energy  of the $O(N)$ singlet CFT.\foot{Ref. \ci{jev}   also   checked this
 matching in the  $d=3$ case  using the  collective  field  formalism.}

 \subsection{Casimir part of the free energy}
\label{Casimir}

 Since  we have   already matched the $N^0$ part of  dual scalar   free energy,
 the Casimir energy part of the  HS  free energy in \rf{20}  should  vanish  after the summation over spins.
 The  Casimir part of the scalar free energy \rf{ca}   scales  as $N$  and thus should  be compared to
 the classical order $N$ part of the HS free energy.

 More precisely, the above  should apply to the standard  HS theory dual to $U(N)$ complex scalar theory.
 In the $O(N)$ real   scalar theory there is a subtlety  noticed   in \ci{gk}:  the matching  should work  provided
 the classical HS coupling constant  is not $N$ but $N-1$   (cf. also \ci{jev}).
 In this case the  Casimir energy of the minimal HS theory should not vanish but should  match the Casimir energy
 of a single real   conformal scalar in $R \times S^{d-1}$, i.e. \rf{a1}.
 In other words,
  we should have the $N\b E_c$ term in the free energy of the boundary theory matching the
  sum  of the classical  $(N-1) \b E_c$  term plus the one-loop  $\b E_c$  term in the minimal   HS  theory.
  We shall  indeed confirm this  below.

 The Casimir  part $F_c =\b E_c$  is the same  as in the  case of the  global AdS$_{d+1}$
 with boundary $R\times S^{d-1}$ (with time interval regularized   by $\b\to \infty$).
 It   is
defined  by the spectrum of the Hamiltonian associated to the global AdS time.
Equivalently, it   can be  also determined  (e.g., via \rf{02},\rf{04})   from  the one-particle  HS  partition functions   \rf{8} and \rf{777}   found
above.
Explicitly,  in the standard HS theory we  should get
\be
&& E_c = \ha \zeta_{E} (-1) \ , \ \ \ \ \ \ \ \ \ \ \ \ \
\z_E (z) ={1 \ov \G(z)} \int^\infty_0 d\b\, \b^{z-1} \   \zz(\b)   
\ ,   \la{35}\\
&& \la{ii}
   \zz(\b)=  { e^{-(d-2)\b } ( 1 + e^{-\b} )^2 \ov (1-e^{-\b})^{2d-2} }= { \cosh^2  { \b \ov 2}\ov  4^{d-2}  \big(\sinh^2{ \b \ov 2}\big)^{d-1}}
\ .    \ee
Using that
\be
(1-q)^{-b} =  \sum_{n=1}^\infty  {\te {n + b -2 \choose b-1} } q^{n-1} \ , \ \ \ \ \ \ \ \ \ \ \ \
{ 1 \ov \G(z) }  \int^\infty_0 d\b\ \b^{z-1}\,   e^{-a \b} = a^{-z} \ , \la{xxx}\ee
this gives
 for a general $d$: \foot{Here (as also  earlier  in \rf{re})  we  have shifted the summation variable and noted that  one can restore  the lower value of the summation  interval
 due to  vanishing  of the coefficients   of the second and third terms   at $n=1,2$.}
\be \la{36}
  \zeta_{E}(z) &=&
   \sum_{n=1}^\infty {\te { n + 2d -4 \choose 2d -3} }\Big[ { 1 \ov ( n + d-3)^z}  + { 2 \ov ( n + d-2)^z}  + { 1 \ov ( n + d-1)^z} \Big]\no \\
    &=&    \sum_{n=1}^\infty { b_n(d)\ov ( n + d-3)^z}  \ , \\
    \la{37}
      b_n(d) &=&  {\te { n + 2d -4 \choose 2d -3} }  +  2 {\te { n + 2d -5 \choose 2d -3} } + {\te { n + 2d -6 \choose 2d -3} }  \no \\
    & =&{\te  { ( n + 2d-6)!    \ov (n-1)!( 2d-3)!}  \Big[ 4 n^2 + 8 ( d -3) n + 4d^2 - 22d + 32 \Big] }
\ .
 \ee
 One can then compute $E_c$ in in \rf{35}, i.e. $E_c = \ha  \sum_{n=1}^\infty { b_n(d)\,  ( n + d-3)} $,
     using the standard  Riemann  $\z$-function regularization, finding that
     the standard  HS theory  vacuum energy in AdS$_{d+1}$ {\it vanishes}   for any $d$,
 \be E_c = \ha \zeta_{E} (-1) =  \sum_{s=0}^\infty   E_{c,s} =0 \ . \la{abo}\ee
 For example,   for the  HS theory in AdS$_4$, AdS$_5$ and AdS$_7$ we get
\be \la{361}d=3: \ \ \
\zeta_{E} (z)  &=& \sum_{n=1}^\infty {\te { 1 \ov 6} }n(n+1) (n+2)\Big[   n^{-z}  +  2  ( n + 1)^{-z}  +  ( n + 2)^{-z} \Big]\no \\
&=& \sum_{n=1}^\infty  {\te { 1 \ov 3}} ( 2n^2 +1)  n^{-z} =    {\te { 1 \ov 3}}\Big[ 2  \z(z-3) + \z(z-1)\Big] \  ,\no
 \\
d=4: \ \ \  \te   \zeta_{E} (z)&=&  \sum_{n=1}^\infty\te  {   { 4 \ov  5! } n (n+1 ) (n+2)  (n^2 + 2n + 2)  ( n + 1)^{-z}}
=  {\te { 1 \ov 30}}\Big[   \z(z-5) - \z(z-1)\Big] \  ,\no   \la{38}\\
d=6: \ \ \   \te  \zeta_{E}(z) & =&  \sum_{n=1}^\infty \te {   { 4  \ov  9!}  n (n+1 ) (n+2)(n+3) (n+4) (n+5) (n+6)    (n^2 + 6n   + 11 ) } ( n + 3)^{-z} \no  \\
&=& {\te { 4 \ov 9!}} \Big[\zeta (z-9)-12 \zeta (z-7)+21 \zeta (z-5)+62 \zeta (z-3)-72 \zeta (z-1)\Big]  \ .
  \la{39}
 \ee
These expressions vanish at
$z=-1$     due to   $ \zeta ( - 2n) =0$.   
Equivalently, one may  use
an  exponential cutoff  $e^{-\ep (n +d-3)}$ with the ``spectral''   parameter $(n +d-3)$
appearing in \rf{36}. Then  the sum in \rf{36}  can be done exactly at $z=-1$, giving for the  regularized vacuum energy \rf{02} (cf. \rf{aaa1})
\be
E_c(\ep)  = \ha \zeta_{E}(-1; \ep) = { 4 e^{-\ep d} \ov  (1 - e^{-\ep} )^{2 d} }
   \Big[ d + (d-2) \cosh \ep \Big]\,  \sinh \ep   \ . \la{366}
 \ee
 Expanding in $\ep\to 0 $  and subtracting the singular $1\ov \ep^k$ terms
  one finds that the  finite part in \rf{366}  is always zero.

 One can  see the reason for  this vanishing  of the  Casimir energy   directly from \rf{35}.   
 Since $\G(-1)= \infty$, the result for $E_c$  can be non-zero only
  if the remaining integral over $\b$  has a pole at $z=-1$. The pole can not appear  since  the partion function
  $\zz(\b)$ appearing in the  integrand  of  \rf{35} is even in $\beta$,
i.e. contains only even powers of $\b$   in its small $\b$ expansion:
$\zz(\b) =  4\b^{-2(d-1)} \big[ 1 + { 1 \ov 12}(d-4) \b^2 + ... \big]$.

\iffa
 a similar argument for the contribution of the minimal theory from the integral in eq. (3.22). In that case the relevant function of q is
$q^(d-2)(1+q^2)/(1-q^2)^{d-1}$
Under $q->1/q (\beta->-\beta)$ this is even if d is odd,  and odd if d is even.  So in odd d the integrand expands in even powers of
$ \beta$ and there is no pole at z=-1, and $\zeta_{min}(-1)$ vanishes due to the $ 1/\Gamma(z) $
factor, while in even d there is a $1/\beta$ pole and the result is non-zero.
\fi

Let us  now repeat the above  analysis  in the  case of the minimal  HS theory with  the one-particle partition function in \rf{777}. Here  we get
 the following analog of \rf{35}
\be
&&E^{\rm min}_c = \ha \zeta^{\rm min}_{E} (-1) \ , \ \ \ \ \ \ \ \ \ \ \ \ \
\z_{E}^{\rm min}(z)  = \ha  \z_E (z)  +   \delta \z (z)  \ , \la{de} \\
&& \delta \z (z) \equiv  \ha {1 \ov \G(z)} \int^\infty_0 d\b\, \b^{z-1} \, { e^{-(d-2)\b } ( 1 + e^{-2\b} ) \ov    (1-e^{-2\b})^{d-1} }
\ , \la{355}
\ee
where $  \z_E (z)$ in \rf{de}  is the standard HS  function given by \rf{35},\rf{36} which vanishes at  $z=-1$ \rf{abo} as discussed above.
Using   \rf{xxx}  we find (cf. \rf{36})
\be
 \delta \z (z) = \ha \sum_{n=1}^\infty  {\te { n + d -3 \choose d -2} } \Big[  ( 2n+  d -4)^{-z}  +    ( 2n+ d -2)^{-z}  \Big]=
 \ha \sum_{n=1}^\infty\te  { (n + d-4)! \ov (d-2)! (n-1)! } ( 2n + d-4)^{1-z}  \ .\ \ \ \ \la{357} \ee
Thus   we  get
\be
E^{\rm min} _c= 
 \ha \delta \z (-1)=  {\te  {1 \ov 4} } \sum_{n=1}^\infty \te { (n + d-4)! \ov (d-2)! (n-1)! } ( 2n + d-4)^{2}=
  \sum_{n=0}^\infty\te  { (n + d-3)! \ov (d-2)!  n! } \big[n + \ha (d-2)\big] ^{2}  \ . \ \ \ \ \la{360}
\ee
Comparing  this   expression to \rf{a1}  we  observe  that it   is exactly the same as the  Casimir energy $E_c$
of a {\it single}   real   conformal scalar   in $R \times S^{d-1}$  given in \rf{1va}.
 We conclude that, as already  mentioned  above,  this  is   consistent with the $N \to N-1$ shift in
identification of the coupling constant in the minimal HS theory -- $O(N)$ real scalar  duality \ci{gk,gks}.

The  equivalence   between the    scalar   Casimir energy   in $R \times S^{d-1}$  \rf{a1}   and the minimal HS  theory  Casimir energy
in AdS$_{d+1}$  \rf{360}
 is seen   at the level of formal  series
so the equality  of the resulting finite expressions  requires  the use of   the same (zeta-function) regularization  on  both
 sides of the duality.

While the Casimir energy  \rf{35}  of  the  standard HS  theory  vanishes in  AdS$_{d+1}$ for any value of $d$,
 the Casimir   energy of the minimal HS theory vanishes only for odd $d$, i.e. in AdS$_{4}$, AdS$_6$, etc.
It is {\it non-vanishing} for even $d$, i.e. in AdS$_{5}$, AdS$_7$, etc. (see  \rf{1va}).
This is  to be compared with the well-known  vanishing of   vacuum energies of  ${\cal N} >4$  extended gauged
supergravities  in AdS$_4$ \ci{ad}  and of  each Kaluza-Klein  level of  the massive spectrum of 11-dimensional
supergravity compactified on $S^7$ \ci{gn,iy}.\foot{The computation of the vacuum  energy of individual fields
in \ci{ad}  still required, of course, the use of the standard zeta-function regularization of the sum over radial quantum number $n$
(as, e.g., in the scalar case on the sphere in \rf{vvv}).}

The expressions for the HS theory  vacuum energies in \rf{abo}  and \rf{360}
  were found  above by first doing the formally convergent sum  over spins $s$  for fixed $\b$ under the integral in
  \rf{35}  and then regularizing the sum over $n$. If instead we first   found the standard  (zeta-function regularized)
  expressions for the Casimir  energies of  each spin   $s$  field   and then summed  over  spin
  we would   get a divergent series  that would require a zeta-function regularization, now of the sum over $s$.
  While   the  cancellation of vacuum energy in supergravities happened  due to
large amount of supersymmetry,   in the HS  theory  it  may be viewed as being   due to a
special (zeta function) definition of the
formally divergent sum over  spins -- a definition that   should be  consistent with the underlying symmetries of HS theory.

To  further  illustrate  the role of the regularization of the sum over spins  (already   emphasized  earlier  in the case  of the
 partition function in the Euclidean AdS$_{d+1}$    with   $S^d$   boundary in \ci{gkpst,gk,gks})
below we shall  consider explicitly the individual
 spin $s$ contributions to the Casimir energy  for  some  particular    values of  dimension $d$.

\def \cN {{\cal N}}

\subsection{Casimir energies  of individual  higher  spin fields  in AdS$_{d+1}$}
\label{indiv-Cas}

The  vacuum   energy  for  a   given massless   spin $s$ field
in  AdS$_{d+1}$   can be found  for  a  general   $d$     by using
the expression for $\zz_s(\b)$ from \rf{3} in the representation \rf{04} for the  corresponding  energy  zeta  function
\be
 \z_{E,s} (z) &=&{1 \ov \G(z)} \int^\infty_0 d\b\, \b^{z-1}  {e^{-(s + d-2)\b } \over (1 - e^{-\b} )^{d} }
 \Big(  \dd_s - \dd_{s-1} e^{-\b} \Big) \no \\
  &=&
  \sum_{n=1}^\infty  {\te { n + d -2 \choose d -1} } \Big[ {\dd_s   ( n+ s + d -3)^{-z} } - {\dd_{s-1}   ( n+ s + d -2)^{-z} } \Big]
 \la{395}\ , \\
E_{c,s} &=&\ha  \z_{E,s} (-1) =    \ha
 \sum_{n=1}^\infty  {\te { n + d -2 \choose d -1} } \Big[ {\dd_s   ( n+ s + d -3) } - {\dd_{s-1}   ( n+ s + d -2) } \Big]
   \ . \la{995}   \ee
The expression for $\dd_s$ was  given in \rf{4}  and we used again the relations in  \rf{xxx}.

For example, for a scalar $s=0$ in \rf{1} (with the operator $-\nabla^2 - 2d +4$)
 we have $\dd_0=1$   and thus\foot{In the scalar case  one
should drop the second  ghost  term  in the bracket in \rf{395} but the  general  expression   \rf{395} applies
also   for $s=0$  case as $\dd_{s-1} ={ ( 2 s + d-4)  (s+ d-4)! \ov (d-2)! (s-1)!} $ vanishes automatically
for $s=0$  if  $d > 3$.} 
\be
&&  E_{c,s=0}= \ha \z_{E,s=0} (-1) =\ha  \sum_{n=0}^\infty\te   { (n + d -1)! \ov  (d -1)!  n!  }    ( n+ d -2) \ , \la{392} \ee
This expression for the vacuum energy of a scalar   (with dimension $\Delta=e_0= d-2$) in AdS$_{d+1}$  is  similar but different
from  the one  \rf{a1} for the   vacuum energy of a conformal scalar    in $R \times S^{d}$ or in  $R\times S^{d-1}$.\foot{For example, for  $d=3$  eq. \rf{392}  gives $E_{c,s=0}=  { 1 \ov 4}   \sum_{n=1}^\infty  ( n + 1)n^2 =  { 1 \ov 4}  \zeta(-3) = { 1 \ov 480}$
while for a  conformal scalar in $R \times S^3$ one finds from \rf{a1}  that $E_{c}={ 1 \ov 2}  \zeta(-3) = { 1 \ov 240}$
and  for $R\times S^2$  one has  $E_{c}=0$.}

Let us   consider cases of few low values of  the boundary dimension $d$.
 Let us start    with $d=2$ or   AdS$_3$   case  ($\dd_s=2, \ s > 1 $)
\be
d=2: \ \
\z_{E,s>1} (z) = 2  \zeta(z, s) \ , \quad
E_{c,s>1}  =    \zeta(-1, s)  =   -\te  {1\ov 12}  \big[  1  +  6 s(s-1) \big] \  , \ \la{d2}
\ee
where  $\zeta(z,a) = \sum^\infty_{r=0} ( r+ a)^{-z}$  is the standard Hurwitz zeta function.
The above formula is applicable for $s\geq 2$ while $s=0,1$ are special cases that need to be discussed separately \cite{gks}.
Note that in $d=2$ case the Casimir energy coefficient is directly related to the conformal anomaly \ci{coc}
but this will  not  be  true in general.

For $d=3$ or AdS$_4$  one  finds
\be
d=3: \ \ \ \  \ \ \ \ \
\z_{E,s} (z) &=& \sum_{n=1}^\infty  \ha {\te  { n (n+1) }} \Big[ (2s+1)    ( n+ s )^{-z}  - (2s-1)   ( n+ s + 1)^{-z}  \Big]\no \\
 &=& \sum_{n=1}^\infty n (n + 2s)  ( n+ s )^{-z}  = \zeta(z-2,s+1) - s^2 \zeta(z, s +1)   \ ,  \la{333}\\
 E_{c,s>0}&=& \ha \zeta_{E,s>0}  (-1)  = { \te {1 \ov 8}  s^4  -  {1 \ov 12}  s^2   + {1 \ov 240} }  \  , \ \ \ \ \ \ \ \ \ \ \ \ \ \
  E_{c,0} = \te {1 \ov 480}  \ . \la{32}
\ee
For completeness,
let us   recall  that the   computation of  the vacuum energies  for massless   higher spin fields
 in AdS$_4$   was originally discussed in
the context of extended gauged supergravities
\ci{ad}, starting directly from the energy  spectrum $E_n=\om_n$  for  massless  spin $s=0, \ha, ..., 2$ fields
(assuming reflective boundary conditions at infinity giving discrete {energy}  spectrum  \ci{Avis:1977yn,Breitenlohner:1982jf}).
Explicitly,
 for  a massless spin $s>0 $   field    corresponding to
 $SO(2,d)=SO(2,3)$  representation  $(e_0,s)=(s+1,s)$  with lowest  energy or dimension  $\Delta=e_0=s+1$
one has\foot{For spin 0 case $
\om_{k,j}=   k+j + e_0 , \     \dd_{k,j} = 2j +1 $   where  $k=0,2,4,...,\ \   j=0, 1,2,
...$  and $e_0=1$ (in conformal coupling case) or $e_0=2$  (in standard massless case).}
 $
\om_{k,j}=   k+j +1, \     \dd_{k,j} = 2j +1 $   where  $k=0,1,2,...$ and $  j=s,
s+1, s+2, ....$. This  leads to the following expression for the corresponding energy
    zeta function   
     \ci{ad}
\be
d=3: \ \ \ \z_{E,s}  (z)  =  \sum_{k=0}^\infty    \sum_{j=s}^\infty  (2j+1) (  k + j + 1 )^{-z}
=  \sum_{r=0}^\infty   (r+1)(r + 2s+1)  ( r + s+1 )^{-z}
 \ . \la{31}
    \ee
    This is  equivalent to \rf{333}  and thus gives  the same expression for the  vacuum energy as in  \rf{32}.

    Note  that  $ E_{c,s>0}$  in \rf{32}  formally applies   also for  $s=0$  if the scalar is  assumed
to be {complex}, i.e. carries the same number (two) of  degrees of freedom as all other massless  spin $s$   fields in $d+1=4$.\foot{One can
check directly that the vacuum energy corresponding to a real
scalar in  either $(1,0)$ or $(2,0)$  representations  is \ci{gn,gpp}
$ E_{c,s=0} = { \te {1 \ov 480} }$.
In general, for
a real scalar  field in representation $(e_0,0)$  one has \ci{gn}

$
\z_{E,s=0}(z;e_0) = \sum_{k,j=0}^\infty (2j+1) ( e_0 + 2 k  + j+1)^{-z}\no =
\ha \Big[ \z(z-2,e_0)   + (3 - 2 e_0) \z(z-1,e_0)   +   (e_0-2) (e_0-1) \z(z,e_0)   \Big]
\ $

so that $\z_{E,s=0}(-1;e_0) =
- { 1 \ov 24} e_0^4 + { 1 \ov 4} e_0^3 - { 1 \ov 2} e_0^2 + { 3 \ov 8} e_0 - { 19 \ov 240} $,
giving
$\ha \zeta_{E,0}(-1; 2) =\ha  \zeta_{E,0}(-1; 1) = {1 \ov 480}$.}
The expression in  \rf{32} is true also   for  half-integer  spins  and thus
  one can directly  apply   \rf{32}  to compute the vacuum energies of extended 4-dimensional
supergravity theories using the  supermultiplet sum rules \ci{cur}
$\sum_{s}  (-1)^{2s}  d(s)  s^p=0$,\  $
p < \cN=1, ..., 8$ where $s=0, {1\ov 2} , 1, {3\ov 2} , 2$
and $d(s)$ are multiplicities  of the spin $s$ fields.
  One then  finds  that  the  vacuum energy vanishes in $\cN>4$ extended gauged supergravities
\ci{ad}.  The  vanishing   of vacuum energy was  found   also at each level of the  massive  KK spectrum of 11-dimensional
supergravity compactified on $S^7$ \ci{gn,iy}.

We  can now  see directly that    similar  cancellation of   vacuum energy   happens  also
 in the  purely bosonic    HS  theory
assuming the sum  over  all  spins  is  zeta-function  regularized  (as  suggested  in \ci{gkpst,gk,gks}):
\be
d=3: \ \ \   \  (E_c)_{\rm HS}= E_{c,0} +   \sum_{s=1}^\infty E_{c,s} = { \te {1 \ov 480} }
+   \sum_{s=1}^\infty  \big( { \te {1 \ov 8}  s^4  -  {1 \ov 12}  s^2   + {1 \ov 240} }\big)  =0 \ , \la{34} \ee
where  we  used   that $\zeta(0) = -\ha, \  \z(-2)=\z(-4) =0$.
Such cancellation    happens also in the minimal HS  theory  in AdS$_4$ where one sums over even spins only.
As discussed above,  this is consistent  (in agreement with AdS/CFT)
with the vanishing \rf{1va} of the  Casimir energy
of a conformal scalar in $d=3$, i.e. in $R \times S^2$.

Let us note   that the  expression  appearing in the  vanishing of  vacuum energy in \rf{34}   is   similar  but not identical
  to the one found  for the vanishing  coefficient of the  UV logarithmically  divergent  part  $( \ln Z_{\rm HS})_{\infty } = - a_{\rm HS} \ln \Lambda$
of  the partition function of  HS theory in the euclidean  AdS$_4$  with $S^3$ as the boundary \ci{gk}
\be
a_{\rm HS}  = { \te {1 \ov 360} }
+   \sum_{s=1}^\infty  \big( { \te {5 \ov 24}  s^4  -  {1 \ov 24}  s^2   + {1 \ov 180} }\big)  =0  \ . \la{379}\ee
In general,  for $d \geq 3$ the spin-dependent  coefficients  in  the Casimir energy and in the  coefficient
of the UV divergence  in AdS$_{d+1}$   appear to be different.

Finally, let  us  consider the AdS$_5$  or $d=4$  case  of  \rf{395},\rf{995}
\be
d=4: \ \
\z_{E,s} (z)&=& \sum_{n=1}^\infty  {\te  { 1\ov 6}}  n (n+1)(n+2) \Big[ (s+1)^2    ( n+ s  +1)^{-z}  -   s^2   ( n+ s + 2)^{-z}  \Big]\no \\
 &=& \sum_{n=1}^\infty { \te { 1 \ov 6}}   n ( n+1)  \Big[ (2s+1) n + 3 s^2 + 4s + 2 \Big]   ( n+ s + 1  )^{-z} \no
 \\
 &=&
 {\te \frac{1}{6}} \Big[   (2 s+1) \zeta( z-3, s+2)
 -  3    s (s+1)  \zeta( z-2, s+2) \no \\
 & &\ \ \  - (2 s+1) \zeta( z-1, s+2)   +  s (s^3+2 s^2+2 s+1) \zeta( z, s+2)\Big]
  \la{334} \ , \\
   E_{c,s} &=& \ha \z_{E,s} (-1) = -{\te  { 1 \ov 1440}}  s(s+1)  \Big[  18  s^2 (s+1)^2 -  14 s (s +1)   -11 \Big] \ . \la{3344}\ee
For example,  for low spin  $s=0,1,2$    fields (with $e_0=s+2 $) this expression  reproduces the  values of the Casimir energies
($ 0, - {11\ov 240}, - {553\ov 240}$) found in \ci{gpp}.
Summing  $E_{c,s}$
over   $s=0,1,2,...$ 
 should  be  done  again  using    an   appropriate  spectral  zeta function regularization  as in \ci{gks}, or, equivalently,
   introducing   a cutoff function $e^{-\epsilon [s + { 1\ov 2} (d-3 ) ]}= e^{-\epsilon (s + { 1 \ov 2} )}$
and dropping all    singular terms in the limit $\ep\to 0$.
 As one readily checks, this   gives
 \be \sum_{s=1}^{\infty} E_{c,s}\,  e^{-\epsilon(s+\frac{1}{2})} \Big|_{\epsilon \to 0, \ \rm fin}  = 0\,.
\label{HS-sum}
\ee
This  is in agreement with our earlier general result
 \rf{abo} obtained  directly from the  total  (summed over spin)
 partition function \rf{ii}  and using the  standard  zeta function regularization of  the sum over $n$ in \rf{36},\rf{abo}.\foot{Note again that the
   6-th order polynomial  in $s$ in \rf{3344}
is  similar  but not equivalent to the   coefficient of  the  logarithmic IR divergence in
  massless  higher spin   partition function in AdS$_5$ \ci{gkpst}
  that also vanishes when summed over  spins with a zeta function regularization \ci{gks}.} Similarly, one can sum up the individual Casimir energies $E_{c,s}$ over even spins only, corresponding to the minimal HS theories in AdS$_{d+1}$, and in all $d$ the result is equal to the Casimir energy of a real conformal scalar on $R\times S^{d-1}$, in agreement with the computation of Section \ref{Casimir}.

\renewcommand{\theequation}{6.\arabic{equation}}
 \setcounter{equation}{0}

\section{Matching fermionic CFT's with higher spin theories}
\label{free-fermions}

Having checked the higher spin  AdS/CFT correspondence for singlet sectors of free scalar field theories on $S^1\times S^{d-1}$, we proceed to   analogous
 checks
for  similar  fermionic theories.
More precisely, we will consider the $U(N)$ singlet sector of the theory of $N$ free Dirac fermions or the $O(N)$ singlet sector
of the theory of $N$ free Majorana fermions on $S^1\times S^{d-1}$
and compare them with appropriately defined higher spin theories in AdS$_{d+1}$.
We will explicitly discuss $d=2,3,4$, but extensions to higher $d$ should not be difficult.
These checks are interesting because the spectra of higher spin currents in such fermionic theories are
more complicated than in the scalar theories. Correspondingly, the dual higher spin description of the fermionic theories generally involves massless gauge
fields in more general representations than the fully symmetric ones \cite{Vasiliev:2004cm,dolan}.

\subsection{$d=2$ }

Let us first discuss $d=2$ fermionic CFT's. We may start with $N$ massless free Dirac fermions and impose the $SU(N)$ singlet condition. This may be accomplished by
gauging the $SU(N)$ symmetry and then adding the WZW term for the $SU(N)$ gauge field $A_\mu = i g \partial_\mu g^{-1}$ with a coefficient $k$.
In the limit $k\rightarrow \infty$ we expect to find the free fermion theory restricted to the $SU(N)$ singlet sector.\footnote{
Such a construction is similar to the Gaberdiel-Gopakumar conjectures \cite{Gaberdiel:2010pz,Gaberdiel:2012uj,Chang:2011mz} which involve coset CFT's in $d=2$.
The $\lambda \rightarrow 0$ limit of the coset CFT used in \cite{Gaberdiel:2010pz,Gaberdiel:2012uj,Chang:2011mz} is simply
the singlet sector of the CFT of $N$ free Dirac fermions, but with
the $U(1)$ current $\bar \psi_i \gamma^\mu \psi^i$ removed \cite{Gaberdiel:2012uj,Datta:2014ska}.
}
Alternatively, we may start with $N$ massless free Majorana fermions and impose the $O(N)$ singlet constraint by similarly gauging the $O(N)$ symmetry.

In $d=2$ CFT's, the Casimir energy on $R\times S^1$ is completely determined by the central charge: $E=-{1\ov 12}c$. Therefore, the AdS$_3$/CFT$_2$ matching of Casimir
energies is equivalent to matching of the central charge $c$. For $U(N)$ and $O(N)$ singlet scalar theories, the central charge matching was carried
out in \cite{gks}. It was found that the sum of higher spin one-loop contributions vanishes in the theory of all integer spins, while in the theory of
even spins it equals 1, which is the central charge of a real scalar field.

Just as in the scalar cases, we find that the $d=2$ $U(N)$ and $O(N)$ singlet fermionic theories contain conserved currents of spin $s>1$, and the dual theories in AdS$_3$ contain corresponding massless gauge fields. The contribution of such fields to the one-loop central charge is as in \rf{d2} \cite{gks}
\be\la{cs1}
c^{(1)}_s =  1 + 6 \, s\, (s - 1) \ ,  \qquad s \geq 2 \,.
\ee
In carrying out the matching for the fermionic theories, we find subtle but important differences from the scalar case, that affect the fields of spin $1$ and $0$.
For the theory of $N$ Dirac fermions, the spin-1 current is $\bar \psi_i \gamma^\mu \psi^i$, and it generates the standard Kac-Moody algebra.
Correspondingly, the dual vector field in AdS$_3$ has the Chern-Simons action \cite{creutzig}. The contribution of the Chern-Simons field
is $ c^{(1)}_{\rm CS}=1$; this can be deduced from the central charge of the current algebra in the dual theory or can be found from a direct calculation in
AdS$_3$.
In contrast, in the $d=2$ scalar CFT the vector current does not satisfy the standard Kac-Moody algebra. It is plausible
to conjecture that in this case
the $s=1$ gauge field in AdS$_3$ has the Maxwell action \cite{gks},
and its contribution to the central charge is $c^{(1)}_{\rm Maxwell}={1 \ov 2} $.
There are two spin-zero operators of dimension
$\Delta=1$ in the fermionic theory: a scalar, $\bar \psi_i \psi^i$ and a pseudoscalar, $\bar\psi_i \gamma_3 \psi^i$. Therefore, the dual theory in AdS$_3$ must contain a complex scalar field with $m^2=-1$ which is right at the BF bound. In general, the contribution of a real
scalar field to one-loop central charge is $c^{(1)}_0 (\Delta)=-\ha (\Delta-1)^3$. Therefore, the complex scalar field makes no contribution to the central charge, while the Chern-Simons vector does, and their total contribution is 1.

Let us compare this with the AdS$_3$ theory dual to the $U(N)$ symmetric scalar model.
Such a theory contains one $m=0$ scalar, which is dual to the $\Delta=0$ operator $\bar \phi_i \phi^i$; it contributes  $c^{(1)}_0=\ha $.
As suggested in \cite{gks}, it also contains a Maxwell field.
Thus, the $s=0,1$ fields in the $U(N)$ singlet fermionic model make the same combined contribution to the central charge as in the scalar model.
The $s>1$ fields work in the same way in the fermionic and scalar models; therefore, the cancelation occurs in both models.

Indeed, in the $U(N)$ invariant fermionic theory, the total one-loop correction to central charge is  $c^{(1)}_{\rm CS}+ \sum_{s=2}^\infty c^{(1)}_s$.
Using the zeta-function regularization for the sum, we see that this vanishes. In the $O(N)$ invariant fermionic theory, the total contribution is the
regularized sum of  $c^{(1)}_s$ over positive even spins. This equals $\ha$, in agreement with the central charge of a single Majorana fermion.
This is consistent with the proposed identification of the coupling constant, $G_N\sim 1/(N-1)$, in the bulk dual of $O(N)$ singlet models.

As a further test of the spectra of the AdS$_3$ theories dual to the $d=2$ fermionic CFT's,
we consider the calculation of the thermal free energy.
 According to \rf{maj}, in $d=2$ one half of the free Dirac  fermion one-particle free  energy is
\footnote{To recall, in general
 dimensions, the character of the free fermion representation of the conformal group is given by \cite{aha,dolan} ${\cal Z}_{1\ov 2} (\beta)
 = \ha Z_{\rm F}(\beta)=n_F \frac{q^{\frac{d-1}{2}}}{(1-q)^{d-1}}$,
 where $n_F=2^{\frac{d}{2}-1}$ for a Weyl fermion in even dimensions, and $n_F=2^{d-1\ov 2}$ for a Dirac fermion in odd dimensions.
 A Dirac fermion in even dimensions can be decomposed as the sum of left and right Weyl spinors.
 }
\begin{equation}
{\cal Z}_{1\ov 2}(\beta)=\frac{2q^{\frac{1}{2}}}{1-q} \ . \la{61}
\end{equation}
According to \rf{24f},
matching of the thermal free energies requires that the square of this partition function
equals the sum of the one-particle partition functions in AdS$_3$.
 Indeed, for the theory dual to the $U(N)$ singlet sector of $N$ free Dirac fermions, we  find
\begin{equation}
\frac{2q}{(1-q)^2}+\sum_{s=1}^{\infty} \Big[\frac{2q^s}{(1-q)^2}-\frac{2q^{s+1}}{(1-q)^2} \Big]
=\frac{4q}{(1-q)^2}=\big[{\cal Z}_{1\ov 2}(\beta)\big]^2\,.
\end{equation}
The first term here is the contribution of the two $\Delta=1$ scalars in AdS$_3$, and the sum corresponds to the contribution of all gauge fields with $s\ge 1$ \cite{creutzig}, including the
Chern-Simons gauge field dual to the spin 1 current $\bar\psi_i \gamma_{\mu}\psi^i$. This matching is essentially a special case of certain identities for product of characters of the conformal group \ci{dolan}.

Now let us consider the minimal higher spin theory in AdS$_3$, which is dual to the $O(N)$ singlet sector of $N$ free Majorana fermions.
Since for Majorana fermions $\bar\psi_i \gamma_{3}\psi^i$ vanishes, this theory has only one real bulk scalar dual to the $\Delta=1$ operator $\bar\psi_i \psi^i$.
It also contains massless gauge fields of positive even spins. Therefore, the sum over one-particle partition functions in AdS$_3$ is
\begin{equation}
\frac{q}{(1-q)^2}+ 2 \sum_{s=2,4,\ldots}^\infty \frac{q^s}{1-q}
=\frac{q+ 3 q^2}{(1-q)^2 (1+q)}=\ha\big[ {\cal Z}_{1\ov 2}(\beta)\big]^2-\ha {\cal Z}_{1\ov 2}(2\beta)\, ,
\end{equation}
and we find agreement with the field theory result (\ref{gf}).

\subsection{$d=3$ }

Next, let us consider the $d=3$ fermionic duality conjectures of \cite{Sezgin:2003pt,Leigh:2003gk}.
In the case of  $N$ free Dirac fermions restricted to the $U(N)$ singlet sector, the ``single trace'' spectrum includes a unique pseudoscalar operator $\bar \psi_i \psi^i$ which has dimension $2$, and a set of totally symmetric higher spin currents, one for each integer spin. The matching of the thermal partition function on $S^1\times S^2$ follows from the identity \cite{dolan, Giombi:2011kc}
\begin{equation}
\frac{q^2}{(1-q)^3}+\sum_{s=1}^{\infty}\Big[(2s+1)\frac{q^{s+1}}{(1-q)^3}-(2s-1)\frac{q^{s+2}}{(1-q)^3}\Big]=\frac{4q^2}{(1-q)^4}=\big[{\cal Z}_{1\ov 2}(\beta)\big]^2\,,
\end{equation}
where ${\cal Z}_{1\ov 2}(\beta)=\frac{2q}{(1-q)^2}$ is  half of the free Dirac fermion one-particle partition function in $d=3$
(see \rf{17f},\rf{maj}). Note that this is the same as the expression (\ref{cri}) for the critical scalar theory.
This is because at large $N$ the spectrum of the free fermion theory is the same as the one of the critical
scalar theory, where the $s=0$ operator $\phi^i\phi^i$ has dimension $\Delta= 2 +{\cal O}(1/N)$, as opposed to
$\Delta=1$ at the free fixed point.\footnote{One difference between the free fermion and critical scalar spectrum is that in the former the $s=0$ operator is a pseudoscalar. However, this difference does not affect the calculation at this order.} Analogously, in the minimal theory with even spins only, one has
\begin{equation}
\frac{q^2}{(1-q)^3}+\sum_{s=2,4,\ldots}^{\infty}\Big[(2s+1)\frac{q^{s+1}}{(1-q)^3}-(2s-1)
\frac{q^{s+2}}{(1-q)^3}\Big]=\ha\big[{\cal Z}_{1\ov 2}(\beta)\big]^2-\ha{\cal Z}_{1\ov 2}(2\beta)\,,
\end{equation}
in agreement with (\ref{gf}).

One may also consider the large $N$ interacting Gross-Neveu model, where the scalar operator has dimension 1 instead of 2. The same $s=0$ operator dimensions appear in the Wilson-Fisher and free scalar models and correspond to the two different
boundary conditions for the $m^2=-2$ scalar in AdS$_4$. As noted in Section \ref{indiv-Cas},
for either choice of scalar
operator dimension, $E_{c,0}={1\ov 480}$. The spectrum of the $s>0$ currents in the $U(N)$ fermionic models (dual to type B Vasiliev theory) is the same as in the $U(N)$ scalar models (dual to type A Vasiliev theory), and
their zeta-function regularized contribution to Casimir energy is $- {1\ov 480}$, properly canceling the $s=0$ contribution.
Similarly, the cancelation of the Casimir energy in the $O(N)$ fermionic models (dual to minimal type B theory) is exactly the same as in the $O(N)$
scalar models (dual to minimal type A theory).

\subsection{$d=4$ }

It is  interesting to look at the higher dimensional free fermion theories,
as in this case  the dual higher   spin theory  contains
new higher spin representations  besides the totally symmetric ones \cite{Vasiliev:2004cm,dolan}.
As an explicit example, let us consider the $d=4$ theory of $N$ free Dirac fermions
restricted to the $U(N)$ singlet sector. The ``single trace"
primary operators in this theory consist of two $\Delta=3$ scalar operators
\begin{equation}
{\cal O} = \bar\psi_i \psi^i \,,\qquad\qquad  \tilde {\cal O} = \bar\psi_i \gamma_5 \psi^i\ ,
\end{equation}
two sets of totally symmetric higher spin currents, schematically \cite{Anselmi:1998bh,Anselmi:1999bb,Alkalaev:2012rg}
\begin{equation}
J_{\mu_1\cdots\mu_s} = \bar\psi_i \gamma_{(\mu_1}\partial_{\mu_2}\cdots \partial_{\mu_s)}\psi^i+\ldots\,,\qquad
\tilde{J}_{\mu_1\cdots\mu_s} = \bar\psi_i \gamma_5 \gamma_{(\mu_1}\partial_{\mu_2}\cdots \partial_{\mu_s)}\psi^i+\ldots\,,\qquad s\ge 1\,,
\end{equation}
and a tower of mixed symmetry higher spin operators of the schematic form
\begin{equation}
B_{\mu_1\cdots \mu_s,\nu} = \bar\psi_i \gamma_{\nu(\mu_1}\partial_{\mu_2}\cdots \partial_{\mu_s)}\psi^i+\ldots\,,\qquad s\ge 1\,.
\la{mixed}
\end{equation}
The latter operators have the symmetries of the Young tableaux with $s$ boxes in the first row and one box in the second row.
This set of primary operators is dual to two AdS$_5$ scalar fields with $m^2=-3$, two towers of totally symmetric higher spin gauge fields, and a tower of mixed symmetry fields corresponding to (\ref{mixed}). In particular, at $s=1$ we have a massive antisymmetric tensor field dual to the operator $\bar\psi_i \gamma_{\mu\nu}\psi^i$. The agreement between ``single trace'' primaries and single particle states in AdS$_5$ can again be seen by computing the thermal partition function on $S^1\times S^3$. Representations of the $d=4$ conformal group can be labelled by $(\Delta;j_1,j_2)$, where $\Delta$ is the conformal dimension and $j_1,j_2$
 the $SU(2)\times SU(2)$ spins. In this notation, the mixed symmetry
  operators (\ref{mixed}) for a given $s$ correspond to the sum of representations
\begin{equation}\te
\Big(s+2;\frac{s+1}{2},\frac{s-1}{2}\Big)\oplus \Big(s+2;\frac{s-1}{2},\frac{s+1}{2}\Big) \ ,
\end{equation}
and the corresponding character, or one-particle partition function, is \cite{dolan}
\begin{equation}
{\cal Z}_{s}^{{\rm mixed}}(\beta) = 2\frac{q^{s+2}}{(1-q)^4}\Big[s(s+2)-q(s^2-1)\Big]\,.
\label{Z-mixed}
\end{equation}
It would be interesting to derive this directly in AdS by computing the heat kernel and one-loop determinant for the mixed symmetry fields in the bulk.
Putting this together with the scalar and totally symmetric higher spin contributions and summing over spins one gets
\begin{eqnarray}
&&\frac{2q^3}{(1-q)^4}+2\sum_{s=1}^{\infty}\frac{q^{s+2}}{(1-q)^4}\Big[(s+1)^2-q s^2\Big]
+2 \sum_{s=1}^{\infty} \frac{q^{s+2}}{(1-q)^4}\Big[s(s+2)-q(s^2-1)\Big]\cr
&&=\frac{16q^3}{(1-q)^6}=\big[{\cal Z}_{1\ov 2}(\beta)\big]^2\,,
\end{eqnarray}
which indeed agrees with \rf{24f} and the form of the free fermion character \rf{17f} in $d=4$.
This is another example of an identity
between a product of characters of two singleton
representations of the conformal group, and the character of the corresponding  direct sum of higher spin representations,
generalizing the Flato-Fronsdal
relation discovered in $d=3$ \ci{Flato:1978qz}.\footnote{In $d=4$  one may consider not only the $j=0$ and $j= \ha$ singletons
\ci{Gunaydin:1998km,Sezgin:2001zs,Vasiliev:2004cm}
(i.e. free massless scalar and spinor fields of boundary  CFT), but also $j=1$ and higher spin ones \ci{Boulanger:2011se}.
In such cases, one finds similar relations between characters \ci{dolan}. The $d=4$ CFT corresponding to the $j=1$ singleton
has $N$ free Maxwell fields restricted to the $O(N)$ singlet sector \cite{Alba:2013yda}. Similar theories may be considered in higher dimensions. For example, in $d=6$,
one may study the $O(N)$ singlet sector of $N$ free anti-symmetric tensor fields; this theory has a higher spin AdS$_7$ dual.}

  By using (\ref{02}), (\ref{04}), the knowledge of ${\cal Z}(\beta)$ for each representation is enough to determine the corresponding Casimir energies of the AdS$_5$ fields. For the $\Delta=3$ scalar field, we get
\begin{equation}
\zeta_{E,0}^{\Delta=3}(z) =\sum_{n=1}^{\infty} \te \frac{1}{6} n (n+1)(n+2)(n+2)^{-z} \ ,
\end{equation}
which yields
\begin{equation}
E_{c,0}^{\Delta=3} =\ha  \zeta_{E,0}^{\Delta=3}(-1)=-\te \frac{1}{480}\,.
\end{equation}
The Casimir energy for the totally symmetric higher spins in AdS$_5$ was already computed in (\ref{334}), (\ref{3344}). Its regularized sum over
all integer spins vanishes \rf{HS-sum}.
For the mixed symmetry fields, from (\ref{Z-mixed}) we get  (using the same regularization of the sum over spins
 as in \rf{HS-sum})
\begin{eqnarray}
&& \zeta_{E,s}^{{\rm mixed}}(z)=\sum_{n=1}^{\infty} \te \frac{1}{6} n (n+1)(n+2)\Big[s(s+2)(n+s+1)^{-z}-(s^2-1)(n+s+2)^{-z}\Big]\ ,\no \ \  \\
&&E_{c,s}^{{\rm mixed}}=\te \frac{1}{2}\zeta_{E,s}^{{\rm mixed}}(-1)=\frac{1}{720}\Big[-3 -19 s(s+1) + 44 s^2(s+1)^2 -18 s^3(s+1)^3\Big]\ , \ \ \\
&&E_{c}^{{\rm mixed}} = \sum_{s=1}^{\infty} \te \frac{1}{720}\Big[-3 -19 s(s+1) +
44 s^2(s+1)^2 -18 s^3(s+1)^3\Big]e^{-\epsilon(s+\frac{1}{2})}\ \Big|_{\epsilon \to 0, \ \rm fin}
= \frac{1}{240}\,.\nonumber\ \
\end{eqnarray}
Thus  the total one-loop bulk Casimir energy is
\begin{eqnarray}
E_c = \te 2\times \big(-\frac{1}{480}\big)+2\times 0+\frac{1}{240} = 0\,.
\end{eqnarray}
This is in agreement with the expected vanishing of order $N^0$ correction  to the Casimir energy
 of $N$ free Dirac fermions. Note that in this section we have chosen to compute the total Casimir energy by summing up the individual Casimir energies of each bulk field with a suitable regulator \cite{gks}. Equivalently, one can obtain the same result by first summing over spins the one-particle partition functions, and then performing the Mellin transform (\ref{04}), as described in Section \ref{Casimir} for the scalar theories.

Analogously, we can consider the theory of $N$ Majorana fermions, restricted to the $O(N)$ singlet sector
as discussed in Section \ref{singlet-fermions}.
 The spectrum of operators is a projection of the one described above for the Dirac case. Given any two Majorana fermions $\chi_1$, $\chi_2$, one has the identities
\begin{eqnarray}
&&\bar\chi_1 \chi_2 = \bar\chi_2\chi_1\,,\qquad \bar\chi_1\gamma_5\chi_2 = \bar\chi_2\gamma_5\chi_1 \ , \\
&&\bar\chi_1 \gamma_{\mu} \chi_2 = -\bar\chi_2\gamma_{\mu}\chi_1\,,\qquad \bar\chi_1 \gamma_{\mu}\gamma_5 \chi_2 = \bar\chi_2\gamma_{\mu}\gamma_5\chi_1\ , \\
&&\bar\chi_1 \gamma_{\mu\nu}\chi_2 = -\bar\chi_2 \gamma_{\mu\nu}\chi_1\,.
\end{eqnarray}
The identities in the first line imply that both $\Delta=3$ scalar operators are present in the Majorana theory. On the other hand, using the identities in the second line and integration by parts, one can see that the totally symmetric operators $J_{\mu_1\cdots\mu_s}$ with odd spins
 and the ``axial" $\tilde{J}_{\mu_1\cdots \mu_s}$ with even spins are projected out (they are total derivatives). This leaves effectively a single tower of totally symmetric higher spins of all integer $s$. Finally, the identity in the last line implies that the mixed symmetry operators $B_{\mu_1\cdots\mu_s,\nu}$ with odd spin are projected out. Then, the total one-loop bulk Casimir energy is
\begin{equation}
E_{c\, {\rm min}} ={\te  2\times \big(-\frac{1}{480}\big)}+0+\sum_{s=2,4,\ldots}^{\infty} E_{c,s}^{\rm{mixed}}\ e^{-\epsilon(s+\frac{1}{2})} \Big|_{\epsilon \to 0, \ \rm fin}  =\te \frac{17}{960}\,.
\end{equation}
This is precisely equal to the Casimir energy of a single Majorana fermion in $d=4$, in agreement with the
 shift $N\rightarrow N-1$ in the HS coupling
  which we observe in the real theories in all dimensions.

 We can also consider the thermal partition function of this free real fermion theory. The sum over one-particle partition functions of the bulk AdS$_5$ fields yields
\begin{eqnarray} \label{fermtherm}
&&\frac{2q^3}{(1-q)^4}+\sum_{s=1}^{\infty}\frac{q^{s+2}}{(1-q)^4}\Big[(s+1)^2-q s^2\Big]
+2 \sum_{s=2,4,\ldots}^{\infty} \frac{q^{s+2}}{(1-q)^4}\Big[s(s+2)-q(s^2-1)\Big]\cr
&&=\frac{8q^3}{(1-q)^6}-\frac{2q^3}{(1-q^2)^3}=\ha \big[{\cal Z}_{1\ov 2}(\beta)\big]^2-\ha{\cal Z}_{1\ov 2}(2\beta)\,.
\end{eqnarray}
This is indeed  in   perfect agreement with  the expression  for the   corresponding
real free fermion partition function on $S^1\times S^3$ with the $O(N)$ singlet constraint
found  in Section \ref{real-scalars}  \rf{gf}.

Finally, let us note that in $d=4$ we could also consider the free theory of $N$ complex Weyl fermions in the $U(N)$ singlet sector. In this theory, the $U(N)$ invariant operators form a single tower of totally symmetric currents with all integer spins $s\ge 1$. In particular, there is no scalar operator and no mixed symmetry operators.\footnote{For a Weyl spinor $\psi_{\alpha}^i$, one can construct a Lorentz scalar by contracting the $SU(2)$ index with $\psi^{\alpha, i}$. However, the corresponding object is not $U(N)$ invariant because $\psi_{\alpha}^i$ and $\psi^{\alpha, i}$ are both in the fundamental of $U(N)$.} The one-loop Casimir energy in the bulk then vanishes due to (\ref{HS-sum}).

\section{Higher spin duals of theories with $N_f$ flavors}
\label{HS-Nf}
It is straightforward to generalize the above calculations to the case of free theories with $N_f$ scalars or fermions in the fundamental representation of $U(N)$ or $O(N)$. To be concrete, let us consider $NN_f$ free complex scalars in the $U(N)$ singlet sector. The spectrum of single trace primaries is then given by
\begin{equation}
{\cal O}^{a}_{\ b}=\bar\phi_{ib} \phi^{ia}\,,\qquad \left(J_{(s)}\right)^{a}_{\ b} \sim  \bar\phi_{ib}\partial^s \phi^{ia}\,,\qquad a,b=1,\ldots,N_f\,.
\end{equation}
The dual HS theory should then be a version of Vasiliev theory where all fields are promoted to matrices carrying the $U(N_f)$ indices \cite{Vasiliev:1999ba}. As usual, the global $U(N_f)$ symmetry of the CFT becomes a $U(N_f)$ gauge symmetry in the bulk. At the free level we simply have $N_f^2$ copies of each field, and the calculations described in Section \ref{scalar-Fbeta} readily lead to the result quoted in (\ref{Nf-scalars}) for the $U(N)$ case.

The situation is slightly more interesting in the $O(N)$ case. For $N_f=1$, recall that all odd spin currents are projected out in this case because the scalar field is real. However, for general $N_f$ it is not difficult to see that there are $N_f(N_f+1)/2$ even spin operators and $N_f(N_f-1)/2$ odd spin ones, corresponding to symmetric or antisymmetric combinations of the flavor indices. Then the sum over the HS one-particle partition functions (\ref{3}) gives
\begin{equation}
\frac{N_f(N_f+1)}{2}\sum_{s=0,2,4,\ldots}{\cal Z}_s(\beta)+\frac{N_f(N_f-1)}{2}\sum_{s=1,3,5,\ldots}{\cal Z}_s(\beta) = \frac{N_f^2}{2}   \big[ \zz_0 (\b)\big]^2   +
\frac{N_f}{2} \zz _0 ( 2\b)\,,
\end{equation}
in agreement with (\ref{Nf-scalars}). Similarly, one can analyze the dual of the fermionic theories with $N_f$ complex or real flavors, and the result for the corresponding higher spin sums is readily seen to agree with (\ref{Nf-fermions}).

Let us also briefly comment on the matching of the Casimir energy. In this case, the CFT predicts that the Casimir term in the thermal free energy should simply be $F_c = N N_f \beta E_c$, with $E_c$ the Casimir energy of a single free field. In the HS dual of the $U(N)$ theories, the results for $N_f=1$ immediately imply that the sum of one-loop Casimir energies vanish. So, assuming that $F_c$ is entirely reproduced by the classical bulk calculation (which we do not address here), we would get a result consistent with the duality. For the $O(N)$ theories, on the other hand, one finds a non-vanishing one-loop Casimir energy. For instance, in the HS dual of the free scalar theories, the sum over one-loop bulk Casimir energies gives, in any dimension\footnote{This result follows from the fact that for $N_f=1$ the sum over all spins vanishes, and the sum over even spins gives the Casimir energy of a single conformal scalar.}
\begin{equation}
\frac{N_f(N_f+1)}{2}\sum_{s=0,2,\ldots}E_{c,s}+\frac{N_f(N_f-1)}{2}\sum_{s=1,3,\ldots} E_{c,s} = N_f E_c^{\rm scalar}\ ,
\end{equation}
where $E_c^{\rm scalar}$ is the Casimir energy of a single conformal scalar. Then, agreement with the duality again requires the same shift $N\rightarrow N-1$ in the map between $N$ and the bulk coupling constant that we observed in the case $N_f=1$, i.e. $G^{-1}_{\rm bulk}\sim N-1$. The same result can be seen to apply to the
$O(N)$ singlet sectors of real fermionic theories for general $N_f$.

\section*{Acknowledgments}
We are  grateful to  J. Dowker, C. Herzog, J. Maldacena,
B. Safdi, H. Schnitzer and E. Skvortsov for  useful  comments.
The work of SG is supported in part by the US NSF under Grant No.~PHY-1318681.
The work of IRK is supported in part by the US NSF under Grant No.~PHY-1314198.
The  work of AAT is supported by the ERC Advanced grant No.290456
``Gauge theory -- string theory duality''
and also by the STFC grant ST/J000353/1.

\bibliographystyle{ssg}
\baselineskip 13pt
\bibliography{Casimir}

\begingroup\raggedright\begin{thebibliography}{100}

\bibitem{Maldacena:1997re}
J.~M. Maldacena, ``{The Large $N$ Limit of Superconformal Field Theories and
  Supergravity},'' {\em Adv. Theor. Math. Phys.} {\bf 2} (1998) 231--252,
  \href{http://xxx.lanl.gov/abs/hep-th/9711200}{{\tt hep-th/9711200}}.

\bibitem{Gubser:1998bc}
S.~S. Gubser, I.~R. Klebanov, and A.~M. Polyakov, ``{Gauge Theory Correlators
  from Non-Critical String Theory},'' {\em Phys. Lett.} {\bf B428} (1998)
  105--114, \href{http://xxx.lanl.gov/abs/hep-th/9802109}{{\tt
  hep-th/9802109}}.

\bibitem{Witten:1998qj}
E.~Witten, ``{Anti-de~Sitter Space and Holography},'' {\em Adv. Theor. Math.
  Phys.} {\bf 2} (1998) 253--291,
  \href{http://xxx.lanl.gov/abs/hep-th/9802150}{{\tt hep-th/9802150}}.

\bibitem{Klebanov:2002ja}
I.~R. Klebanov and A.~M. Polyakov, ``{AdS dual of the critical $O(N)$ vector
  model},'' {\em Phys. Lett.} {\bf B550} (2002) 213--219,
  \href{http://xxx.lanl.gov/abs/hep-th/0210114}{{\tt hep-th/0210114}}.

\bibitem{Sundborg:2000wp}
B.~Sundborg, ``{Stringy gravity, interacting tensionless strings and massless
  higher spins},'' {\em Nucl.Phys.Proc.Suppl.} {\bf 102} (2001) 113--119,
  \href{http://xxx.lanl.gov/abs/hep-th/0103247}{{\tt hep-th/0103247}}.

\bibitem{Fradkin:1987ks}
E.~Fradkin and M.~A. Vasiliev, ``{On the Gravitational Interaction of Massless
  Higher Spin Fields},'' {\em Phys.Lett.} {\bf B189} (1987) 89--95.

\bibitem{Vasiliev:1990en}
M.~A. Vasiliev, ``{Consistent equation for interacting gauge fields of all
  spins in (3+1)-dimensions},'' {\em Phys.Lett.} {\bf B243} (1990) 378--382.

\bibitem{Vasiliev:1992av}
M.~A. Vasiliev, ``{More on equations of motion for interacting massless fields
  of all spins in (3+1)-dimensions},'' {\em Phys. Lett.} {\bf B285} (1992)
  225--234.

\bibitem{Vasiliev:1995dn}
M.~A. Vasiliev, ``{Higher-spin gauge theories in four, three and two
  dimensions},'' {\em Int. J. Mod. Phys.} {\bf D5} (1996) 763--797,
  \href{http://xxx.lanl.gov/abs/hep-th/9611024}{{\tt hep-th/9611024}}.

\bibitem{Prokushkin:1998bq}
S.~Prokushkin and M.~A. Vasiliev, ``{Higher spin gauge interactions for massive
  matter fields in 3-D AdS space-time},'' {\em Nucl.Phys.} {\bf B545} (1999)
  385, \href{http://xxx.lanl.gov/abs/hep-th/9806236}{{\tt hep-th/9806236}}.

\bibitem{Vasiliev:1999ba}
M.~A. Vasiliev, ``{Higher spin gauge theories: Star-product and AdS space. In:
  Shifman, M.A. (ed.): The many faces of the superworld, 533-610 },''
  \href{http://xxx.lanl.gov/abs/hep-th/9910096}{{\tt hep-th/9910096}}.

\bibitem{Vasiliev:2003ev}
M.~Vasiliev, ``{Nonlinear equations for symmetric massless higher spin fields
  in (A)dS(d)},'' {\em Phys.Lett.} {\bf B567} (2003) 139--151,
  \href{http://xxx.lanl.gov/abs/hep-th/0304049}{{\tt hep-th/0304049}}.

\bibitem{Bekaert:2005vh}
X.~Bekaert, S.~Cnockaert, C.~Iazeolla, and M.~Vasiliev, ``{Nonlinear higher
  spin theories in various dimensions},''
  \href{http://xxx.lanl.gov/abs/hep-th/0503128}{{\tt hep-th/0503128}}.

\bibitem{Sezgin:2003pt}
E.~Sezgin and P.~Sundell, ``{Holography in 4D (super) higher spin theories and
  a test via cubic scalar couplings},'' {\em JHEP} {\bf 07} (2005) 044,
  \href{http://xxx.lanl.gov/abs/hep-th/0305040}{{\tt hep-th/0305040}}.

\bibitem{Leigh:2003gk}
R.~G. Leigh and A.~C. Petkou, ``{Holography of the N = 1 higher-spin theory on
  AdS(4)},'' {\em JHEP} {\bf 06} (2003) 011,
  \href{http://xxx.lanl.gov/abs/hep-th/0304217}{{\tt hep-th/0304217}}.

\bibitem{sy}
S.~H. Shenker and X.~Yin, ``{Vector Models in the Singlet Sector at Finite
  Temperature},'' \href{http://xxx.lanl.gov/abs/1109.3519}{{\tt 1109.3519}}.

\bibitem{Sundborg:1999ue}
B.~Sundborg, ``{The Hagedorn transition, deconfinement and N=4 SYM theory},''
  {\em Nucl.Phys.} {\bf B573} (2000) 349--363,
  \href{http://xxx.lanl.gov/abs/hep-th/9908001}{{\tt hep-th/9908001}}.

\bibitem{aha}
O.~Aharony, J.~Marsano, S.~Minwalla, K.~Papadodimas, and M.~Van~Raamsdonk,
  ``{The Hagedorn - deconfinement phase transition in weakly coupled large N
  gauge theories},'' {\em Adv.Theor.Math.Phys.} {\bf 8} (2004) 603--696,
  \href{http://xxx.lanl.gov/abs/hep-th/0310285}{{\tt hep-th/0310285}}.

\bibitem{sh}
H.~J. Schnitzer, ``{Confinement/deconfinement transition of large N gauge
  theories with N(f) fundamentals: N(f)/N finite},'' {\em Nucl.Phys.} {\bf
  B695} (2004) 267--282, \href{http://xxx.lanl.gov/abs/hep-th/0402219}{{\tt
  hep-th/0402219}}.

\bibitem{Flato:1978qz}
M.~Flato and C.~Fronsdal, ``{One Massless Particle Equals Two Dirac Singletons:
  Elementary Particles in a Curved Space. 6.},'' {\em Lett.Math.Phys.} {\bf 2}
  (1978) 421--426.

\bibitem{Kutasov:2000td}
D.~Kutasov and F.~Larsen, ``{Partition sums and entropy bounds in weakly
  coupled CFT},'' {\em JHEP} {\bf 0101} (2001) 001,
  \href{http://xxx.lanl.gov/abs/hep-th/0009244}{{\tt hep-th/0009244}}.

\bibitem{Polyakov:2001af}
A.~M. Polyakov, ``{Gauge fields and space-time},'' {\em Int.J.Mod.Phys.} {\bf
  A17S1} (2002) 119--136, \href{http://xxx.lanl.gov/abs/hep-th/0110196}{{\tt
  hep-th/0110196}}.

\bibitem{bia1}
M.~Bianchi, J.~F. Morales, and H.~Samtleben, ``{On stringy AdS(5) x S5 and
  higher spin holography},'' {\em JHEP} {\bf 0307} (2003) 062,
  \href{http://xxx.lanl.gov/abs/hep-th/0305052}{{\tt hep-th/0305052}}.

\bibitem{bia2}
N.~Beisert, M.~Bianchi, J.~Morales, and H.~Samtleben, ``{On the spectrum of AdS
  / CFT beyond supergravity},'' {\em JHEP} {\bf 0402} (2004) 001,
  \href{http://xxx.lanl.gov/abs/hep-th/0310292}{{\tt hep-th/0310292}}.

\bibitem{Vasiliev:2004cm}
M.~Vasiliev, ``{Higher spin superalgebras in any dimension and their
  representations},'' {\em JHEP} {\bf 0412} (2004) 046,
  \href{http://xxx.lanl.gov/abs/hep-th/0404124}{{\tt hep-th/0404124}}.

\bibitem{Barabanschikov:2005ri}
A.~Barabanschikov, L.~Grant, L.~L. Huang, and S.~Raju, ``{The Spectrum of Yang
  Mills on a sphere},'' {\em JHEP} {\bf 0601} (2006) 160,
  \href{http://xxx.lanl.gov/abs/hep-th/0501063}{{\tt hep-th/0501063}}.

\bibitem{dolan}
F.~Dolan, ``{Character formulae and partition functions in higher dimensional
  conformal field theory},'' {\em J.Math.Phys.} {\bf 47} (2006) 062303,
  \href{http://xxx.lanl.gov/abs/hep-th/0508031}{{\tt hep-th/0508031}}.

\bibitem{jev}
A.~Jevicki, K.~Jin, and J.~Yoon, ``{1/N and Loop Corrections in Higher Spin
  AdS$_4$/CFT$_3$ Duality},'' \href{http://xxx.lanl.gov/abs/1401.3318}{{\tt
  1401.3318}}.

\bibitem{Das:2003vw}
S.~R. Das and A.~Jevicki, ``{Large N collective fields and holography},'' {\em
  Phys.Rev.} {\bf D68} (2003) 044011,
  \href{http://xxx.lanl.gov/abs/hep-th/0304093}{{\tt hep-th/0304093}}.

\bibitem{Giombi:2009wh}
S.~Giombi and X.~Yin, ``{Higher Spin Gauge Theory and Holography: The
  Three-Point Functions},'' {\em JHEP} {\bf 1009} (2010) 115,
  \href{http://xxx.lanl.gov/abs/0912.3462}{{\tt 0912.3462}}.

\bibitem{Giombi:2010vg}
S.~Giombi and X.~Yin, ``{Higher Spins in AdS and Twistorial Holography},'' {\em
  JHEP} {\bf 1104} (2011) 086, \href{http://xxx.lanl.gov/abs/1004.3736}{{\tt
  1004.3736}}.

\bibitem{Giombi:2011ya}
S.~Giombi and X.~Yin, ``{On Higher Spin Gauge Theory and the Critical O(N)
  Model},'' {\em Phys.Rev.} {\bf D85} (2012) 086005,
  \href{http://xxx.lanl.gov/abs/1105.4011}{{\tt 1105.4011}}.

\bibitem{Maldacena:2011jn}
J.~Maldacena and A.~Zhiboedov, ``{Constraining Conformal Field Theories with A
  Higher Spin Symmetry},'' {\em J.Phys.} {\bf A46} (2013) 214011,
  \href{http://xxx.lanl.gov/abs/1112.1016}{{\tt 1112.1016}}.

\bibitem{Maldacena:2012sf}
J.~Maldacena and A.~Zhiboedov, ``{Constraining conformal field theories with a
  slightly broken higher spin symmetry},'' {\em Class.Quant.Grav.} {\bf 30}
  (2013) 104003, \href{http://xxx.lanl.gov/abs/1204.3882}{{\tt 1204.3882}}.

\bibitem{Colombo:2012jx}
N.~Colombo and P.~Sundell, ``{Higher Spin Gravity Amplitudes From Zero-form
  Charges},'' \href{http://xxx.lanl.gov/abs/1208.3880}{{\tt 1208.3880}}.

\bibitem{Didenko:2012tv}
V.~Didenko and E.~Skvortsov, ``{Exact higher-spin symmetry in CFT: all
  correlators in unbroken Vasiliev theory},'' {\em JHEP} {\bf 1304} (2013) 158,
  \href{http://xxx.lanl.gov/abs/1210.7963}{{\tt 1210.7963}}.

\bibitem{Gelfond:2013xt}
O.~Gelfond and M.~Vasiliev, ``{Operator algebra of free conformal currents via
  twistors},'' {\em Nucl.Phys.} {\bf B876} (2013) 871--917,
  \href{http://xxx.lanl.gov/abs/1301.3123}{{\tt 1301.3123}}.

\bibitem{Didenko:2013bj}
V.~Didenko, J.~Mei, and E.~Skvortsov, ``{Exact higher-spin symmetry in CFT:
  free fermion correlators from Vasiliev Theory},'' {\em Phys.Rev.} {\bf D88}
  (2013) 046011, \href{http://xxx.lanl.gov/abs/1301.4166}{{\tt 1301.4166}}.

\bibitem{gk}
S.~Giombi and I.~R. Klebanov, ``{One Loop Tests of Higher Spin AdS/CFT},'' {\em
  JHEP} {\bf 1312} (2013) 068, \href{http://xxx.lanl.gov/abs/1308.2337}{{\tt
  1308.2337}}.

\bibitem{Didenko:2012vh}
V.~Didenko and E.~Skvortsov, ``{Towards higher-spin holography in ambient space
  of any dimension},'' {\em J.Phys.} {\bf A46} (2013) 214010,
  \href{http://xxx.lanl.gov/abs/1207.6786}{{\tt 1207.6786}}.

\bibitem{gks}
S.~Giombi, I.~R. Klebanov, and B.~R. Safdi, ``{Higher Spin AdS$_{d+1}$/CFT$_d$
  at One Loop},'' \href{http://xxx.lanl.gov/abs/1401.0825}{{\tt 1401.0825}}.

\bibitem{Giombi:2011kc}
S.~Giombi, S.~Minwalla, S.~Prakash, S.~P. Trivedi, S.~R. Wadia, {\em et.~al.},
  ``{Chern-Simons Theory with Vector Fermion Matter},'' {\em Eur.Phys.J.} {\bf
  C72} (2012) 2112, \href{http://xxx.lanl.gov/abs/1110.4386}{{\tt 1110.4386}}.

\bibitem{Aharony:2011jz}
O.~Aharony, G.~Gur-Ari, and R.~Yacoby, ``{d=3 Bosonic Vector Models Coupled to
  Chern-Simons Gauge Theories},'' {\em JHEP} {\bf 1203} (2012) 037,
  \href{http://xxx.lanl.gov/abs/1110.4382}{{\tt 1110.4382}}.

\bibitem{ad}
B.~Allen and S.~Davis, ``{Vacuum Energy in Gauged Extended Supergravity},''
  {\em Phys.Lett.} {\bf B124} (1983) 353.

\bibitem{gn}
G.~Gibbons and H.~Nicolai, ``{One Loop Effects on the Round Seven Sphere},''
  {\em Phys.Lett.} {\bf B143} (1984) 108--114.

\bibitem{iy}
T.~Inami and K.~Yamagishi, ``{Vanishing Quantum Vacuum Energy in
  Eleven-dimensional Supergravity on the Round Seven Sphere},'' {\em
  Phys.Lett.} {\bf B143} (1984) 115--120.

\bibitem{Balasubramanian:1999re}
V.~Balasubramanian and P.~Kraus, ``{A Stress tensor for Anti-de Sitter
  gravity},'' {\em Commun.Math.Phys.} {\bf 208} (1999) 413--428,
  \href{http://xxx.lanl.gov/abs/hep-th/9902121}{{\tt hep-th/9902121}}.

\bibitem{herz}
C.~P. Herzog and K.-W. Huang, ``{Stress Tensors from Trace Anomalies in
  Conformal Field Theories},'' {\em Phys.Rev.} {\bf D87} (2013) 081901,
  \href{http://xxx.lanl.gov/abs/1301.5002}{{\tt 1301.5002}}.

\bibitem{hua}
K.-W. Huang, ``{Weyl Anomaly Induced Stress Tensors in General Manifolds},''
  {\em Nucl.Phys.} {\bf B879} (2014) 370--381,
  \href{http://xxx.lanl.gov/abs/1308.2355}{{\tt 1308.2355}}.

\bibitem{Henningson:1998gx}
M.~Henningson and K.~Skenderis, ``{The Holographic Weyl anomaly},'' {\em JHEP}
  {\bf 9807} (1998) 023, \href{http://xxx.lanl.gov/abs/hep-th/9806087}{{\tt
  hep-th/9806087}}.

\bibitem{Mansfield:2002pa}
P.~Mansfield, D.~Nolland, and T.~Ueno, ``{Order $1 / N^2$ test of the Maldacena
  conjecture. 2. The Full bulk one loop contribution to the boundary Weyl
  anomaly},'' {\em Phys.Lett.} {\bf B565} (2003) 207--210,
  \href{http://xxx.lanl.gov/abs/hep-th/0208135}{{\tt hep-th/0208135}}.

\bibitem{Ardehali:2013gra}
A.~Arabi~Ardehali, J.~T. Liu, and P.~Szepietowski, ``{The spectrum of IIB
  supergravity on $AdS_5$ x $S^5/Z_3$ and a $1/N^2$ test of AdS/CFT},'' {\em
  JHEP} {\bf 1306} (2013) 024, \href{http://xxx.lanl.gov/abs/1304.1540}{{\tt
  1304.1540}}.

\bibitem{Ardehali:2013xya}
A.~A. Ardehali, J.~T. Liu, and P.~Szepietowski, ``{$1/N^2$ corrections to the
  holographic Weyl anomaly},'' {\em JHEP} {\bf 1401} (2014) 002,
  \href{http://xxx.lanl.gov/abs/1310.2611}{{\tt 1310.2611}}.

\bibitem{Anselmi:1998bh}
D.~Anselmi, ``{Theory of higher spin tensor currents and central charges},''
  {\em Nucl.Phys.} {\bf B541} (1999) 323--368,
  \href{http://xxx.lanl.gov/abs/hep-th/9808004}{{\tt hep-th/9808004}}.

\bibitem{Anselmi:1999bb}
D.~Anselmi, ``{Higher spin current multiplets in operator product
  expansions},'' {\em Class.Quant.Grav.} {\bf 17} (2000) 1383--1400,
  \href{http://xxx.lanl.gov/abs/hep-th/9906167}{{\tt hep-th/9906167}}.

\bibitem{Alkalaev:2012rg}
K.~Alkalaev, ``{Mixed-symmetry tensor conserved currents and AdS/CFT
  correspondence},'' {\em J.Phys.} {\bf A46} (2013) 214007,
  \href{http://xxx.lanl.gov/abs/1207.1079}{{\tt 1207.1079}}.

\bibitem{gpp}
G.~Gibbons, M.~Perry, and C.~Pope, ``{Partition functions, the Bekenstein bound
  and temperature inversion in anti-de Sitter space and its conformal
  boundary},'' {\em Phys.Rev.} {\bf D74} (2006) 084009,
  \href{http://xxx.lanl.gov/abs/hep-th/0606186}{{\tt hep-th/0606186}}.

\bibitem{Dowker:1978md}
J.~Dowker and G.~Kennedy, ``{Finite Temperature and Boundary Effects in Static
  Space-Times},'' {\em J.Phys.} {\bf A11} (1978) 895.

\bibitem{Dowker:1983ci}
J.~Dowker, ``{Finite Temperature and Vacuum Effects in Higher Dimensions},''
  {\em Class.Quant.Grav.} {\bf 1} (1984) 359.

\bibitem{Blau:1988kv}
S.~Blau, M.~Visser, and A.~Wipf, ``{Zeta Functions and the Casimir Energy},''
  {\em Nucl.Phys.} {\bf B310} (1988) 163,
  \href{http://xxx.lanl.gov/abs/0906.2817}{{\tt 0906.2817}}.

\bibitem{coc}
A.~Cappelli and A.~Coste, ``{On the Stress Tensor of Conformal Field Theories
  in Higher Dimensions},'' {\em Nucl.Phys.} {\bf B314} (1989) 707.

\bibitem{oz1}
M.~Ozcan, ``{Casimir energy density for spherical universes in n-dimensional
  spacetime},'' {\em Class.Quant.Grav.} {\bf 23} (2006) 5531--5546.

\bibitem{oz2}
M.~Ozcan, ``{Green's function for a n-dimensional closed, static universe and
  with a spherical boundary},''
  \href{http://xxx.lanl.gov/abs/gr-qc/0106082}{{\tt gr-qc/0106082}}.

\bibitem{byt}
A.~A. Bytsenko, G.~Cognola, L.~Vanzo, and S.~Zerbini, ``{Quantum fields and
  extended objects in space-times with constant curvature spatial section},''
  {\em Phys.Rept.} {\bf 266} (1996) 1--126,
  \href{http://xxx.lanl.gov/abs/hep-th/9505061}{{\tt hep-th/9505061}}.

\bibitem{Allen:1986qi}
B.~Allen, ``{Does statistical mechanics equal one loop quantum field
  theory?},'' {\em Phys.Rev.} {\bf D33} (1986) 3640.

\bibitem{Fursaev:2001yu}
D.~V. Fursaev, ``{Statistical mechanics, gravity, and Euclidean theory},'' {\em
  Nucl.Phys.Proc.Suppl.} {\bf 104} (2002) 33--62,
  \href{http://xxx.lanl.gov/abs/hep-th/0107089}{{\tt hep-th/0107089}}.

\bibitem{Chang:2012kt}
C.-M. Chang, S.~Minwalla, T.~Sharma, and X.~Yin, ``{ABJ Triality: from Higher
  Spin Fields to Strings},'' {\em J.Phys.} {\bf A46} (2013) 214009,
  \href{http://xxx.lanl.gov/abs/1207.4485}{{\tt 1207.4485}}.

\bibitem{meh}
M.~Mehta, ``{Random Matrices},'' {\em Academic Press, N.Y.} (1967).

\bibitem{Sinha:2000ap}
S.~Sinha and C.~Vafa, ``{SO and Sp Chern-Simons at large N},''
  \href{http://xxx.lanl.gov/abs/hep-th/0012136}{{\tt hep-th/0012136}}.

\bibitem{Fronsdal:1978vb}
C.~Fronsdal, ``{Singletons and Massless, Integral Spin Fields on de Sitter
  Space (Elementary Particles in a Curved Space. 7.},'' {\em Phys.Rev.} {\bf
  D20} (1979) 848--856.

\bibitem{Metsaev:1994ys}
R.~Metsaev, ``{Lowest eigenvalues of the energy operator for totally
  (anti)symmetric massless fields of the n-dimensional anti-de Sitter group},''
  {\em Class.Quant.Grav.} {\bf 11} (1994) L141--L145.

\bibitem{Metsaev:1997nj}
R.~Metsaev, ``{Arbitrary spin massless bosonic fields in d-dimensional anti-de
  Sitter space},'' {\em Lect.Notes Phys.} {\bf 524} (1999) 331--340,
  \href{http://xxx.lanl.gov/abs/hep-th/9810231}{{\tt hep-th/9810231}}.

\bibitem{gl}
R.~K. Gupta and S.~Lal, ``{Partition Functions for Higher-Spin theories in
  AdS},'' {\em JHEP} {\bf 1207} (2012) 071,
  \href{http://xxx.lanl.gov/abs/1205.1130}{{\tt 1205.1130}}.

\bibitem{Gaberdiel:2010ar}
M.~R. Gaberdiel, R.~Gopakumar, and A.~Saha, ``{Quantum $W$-symmetry in
  $AdS_3$},'' {\em JHEP} {\bf 1102} (2011) 004,
  \href{http://xxx.lanl.gov/abs/1009.6087}{{\tt 1009.6087}}.

\bibitem{Avis:1977yn}
S.~Avis, C.~Isham, and D.~Storey, ``{Quantum Field Theory in anti-De Sitter
  Space-Time},'' {\em Phys.Rev.} {\bf D18} (1978) 3565.

\bibitem{Breitenlohner:1982jf}
P.~Breitenlohner and D.~Z. Freedman, ``{Stability in Gauged Extended
  Supergravity},'' {\em Annals Phys.} {\bf 144} (1982) 249.

\bibitem{Camporesi:1991nw}
R.~Camporesi, ``{zeta function regularization of one loop effective potentials
  in anti-de Sitter space-time},'' {\em Phys.Rev.} {\bf D43} (1991) 3958--3965.

\bibitem{Camporesi:1992wn}
R.~Camporesi and A.~Higuchi, ``{Stress energy tensors in anti-de Sitter
  space-time},'' {\em Phys.Rev.} {\bf D45} (1992) 3591--3603.

\bibitem{Camporesi:1993mz}
R.~Camporesi and A.~Higuchi, ``{Arbitrary spin effective potentials in anti-de
  Sitter space-time},'' {\em Phys.Rev.} {\bf D47} (1993) 3339--3344.

\bibitem{Giombi:2008vd}
S.~Giombi, A.~Maloney, and X.~Yin, ``{One-loop Partition Functions of 3D
  Gravity},'' {\em JHEP} {\bf 0808} (2008) 007,
  \href{http://xxx.lanl.gov/abs/0804.1773}{{\tt 0804.1773}}.

\bibitem{David:2009xg}
J.~David, M.~R. Gaberdiel, and R.~Gopakumar, ``{The Heat Kernel on AdS(3) and
  its Applications},'' {\em JHEP} {\bf 1004} (2010) 125,
  \href{http://xxx.lanl.gov/abs/0911.5085}{{\tt 0911.5085}}.

\bibitem{ggl}
R.~Gopakumar, R.~K. Gupta, and S.~Lal, ``{The Heat Kernel on $AdS$},'' {\em
  JHEP} {\bf 1111} (2011) 010, \href{http://xxx.lanl.gov/abs/1103.3627}{{\tt
  1103.3627}}.

\bibitem{ke}
C.~Keeler and G.~S. Ng, ``{Partition Functions in Even Dimensional AdS via
  Quasinormal Mode Methods},'' \href{http://xxx.lanl.gov/abs/1401.7016}{{\tt
  1401.7016}}.

\bibitem{Gibbons:2011sg}
G.~Gibbons, ``{Anti-de-Sitter spacetime and its uses},''
  \href{http://xxx.lanl.gov/abs/1110.1206}{{\tt 1110.1206}}.

\bibitem{Dowker:1983nt}
J.~Dowker, ``{Arbitrary Spin Theory in the Einstein Universe},'' {\em
  Phys.Rev.} {\bf D28} (1983) 3013.

\bibitem{Allen:1986ty}
B.~Allen, A.~Folacci, and G.~Gibbons, ``{Anti-DeSitter space at finite
  temperature},'' {\em Phys.Lett.} {\bf B189} (1987) 304.

\bibitem{Sakai:1984fg}
N.~Sakai and Y.~Tanii, ``{Effective Potential in Two-dimensional Anti-de Sitter
  Space},'' {\em Nucl.Phys.} {\bf B255} (1985) 401.

\bibitem{my}
E.~Myers, ``{On the Interpretation of the Energy of the Vacuum as the Sum Over
  Zero Point Energies},'' {\em Phys.Rev.Lett.} {\bf 59} (1987) 165.

\bibitem{Dowker:1989gp}
J.~Dowker and J.~P. Schofield, ``{Chemical Potentials in Curved Space},'' {\em
  Nucl.Phys.} {\bf B327} (1989) 267.

\bibitem{Kirsten:1991mi}
K.~Kirsten, ``{Grand thermodynamic potential in a static space-time with
  boundary},'' {\em Class.Quant.Grav.} {\bf 8} (1991) 2239--2255.

\bibitem{Dowker:1989gw}
J.~Dowker, ``{Conformal Properties of the Heat-kernel Expansion: Application to
  the Effective Lagrangian},'' {\em Phys.Rev.} {\bf D39} (1989) 1235--1238.

\bibitem{gkpst}
S.~Giombi, I.~R. Klebanov, S.~S. Pufu, B.~R. Safdi, and G.~Tarnopolsky, ``{AdS
  Description of Induced Higher-Spin Gauge Theory},'' {\em JHEP} {\bf 1310}
  (2013) 016, \href{http://xxx.lanl.gov/abs/1306.5242}{{\tt 1306.5242}}.

\bibitem{cur}
T.~L. Curtright, ``{Charge Renormalization and High Spin Fields},'' {\em
  Phys.Lett.} {\bf B102} (1981) 17.

\bibitem{Gaberdiel:2010pz}
M.~R. Gaberdiel and R.~Gopakumar, ``{An $AdS_3$ Dual for Minimal Model CFTs},''
  {\em Phys.Rev.} {\bf D83} (2011) 066007,
  \href{http://xxx.lanl.gov/abs/1011.2986}{{\tt 1011.2986}}.

\bibitem{Gaberdiel:2012uj}
M.~R. Gaberdiel and R.~Gopakumar, ``{Minimal Model Holography},'' {\em J.Phys.}
  {\bf A46} (2013) 214002, \href{http://xxx.lanl.gov/abs/1207.6697}{{\tt
  1207.6697}}.

\bibitem{Chang:2011mz}
C.-M. Chang and X.~Yin, ``{Higher Spin Gravity with Matter in $AdS_3$ and Its
  CFT Dual},'' {\em JHEP} {\bf 1210} (2012) 024,
  \href{http://xxx.lanl.gov/abs/1106.2580}{{\tt 1106.2580}}.

\bibitem{Datta:2014ska}
S.~Datta, J.~R. David, M.~Ferlaino, and S.~P. Kumar, ``{Higher spin
  entanglement entropy from CFT},''
  \href{http://xxx.lanl.gov/abs/1402.0007}{{\tt 1402.0007}}.

\bibitem{creutzig}
T.~Creutzig, Y.~Hikida, and P.~B. Ronne, ``{Higher spin AdS$_3$ supergravity
  and its dual CFT},'' {\em JHEP} {\bf 1202} (2012) 109,
  \href{http://xxx.lanl.gov/abs/1111.2139}{{\tt 1111.2139}}.

\bibitem{Gunaydin:1998km}
M.~Gunaydin and D.~Minic, ``{Singletons, doubletons and M theory},'' {\em
  Nucl.Phys.} {\bf B523} (1998) 145--157,
  \href{http://xxx.lanl.gov/abs/hep-th/9802047}{{\tt hep-th/9802047}}.

\bibitem{Sezgin:2001zs}
E.~Sezgin and P.~Sundell, ``{Doubletons and 5-D higher spin gauge theory},''
  {\em JHEP} {\bf 0109} (2001) 036,
  \href{http://xxx.lanl.gov/abs/hep-th/0105001}{{\tt hep-th/0105001}}.

\bibitem{Boulanger:2011se}
N.~Boulanger and E.~Skvortsov, ``{Higher-spin algebras and cubic interactions
  for simple mixed-symmetry fields in AdS spacetime},'' {\em JHEP} {\bf 1109}
  (2011) 063, \href{http://xxx.lanl.gov/abs/1107.5028}{{\tt 1107.5028}}.

\bibitem{Alba:2013yda}
V.~Alba and K.~Diab, ``{Constraining conformal field theories with a higher
  spin symmetry in d=4},'' \href{http://xxx.lanl.gov/abs/1307.8092}{{\tt
  1307.8092}}.

\end{thebibliography}\endgroup

\end{document}